\documentstyle[12pt,epsfig,citesort]{article}
\newcommand{\Z}{{\sf Z \!\!\! Z}}

\newcommand{\1}{{\sf 1 \!\! 1}}

\newcommand{\p}{\partial}

\newcommand{\ren}{\mathrm{ren}}
\makeatletter
\@addtoreset{equation}{section}
\makeatother

\title{Nonlinear Realization of Chiral Symmetry \\ on the Lattice}

\author{S.~Chandrasekharan$^{a}$, 
M.~Pepe$^b$, 
F.~D.~Steffen$^b$ $\!\!\!$
\footnote{Present address:
DESY Theory Group, 
Notkestrasse 85, 
D-22603 Hamburg, 
Germany}\, , 
and \\
U.-J.~Wiese$^b$ $\!\!\!$
\footnote{On leave from MIT.}
\\ \\
$^a$ Department of Physics, Duke University, P.O.~Box 90305\\
Durham, North Carolina 27708-0305, USA \\
$^b$ Institute for Theoretical Physics, Bern University \\
Sidlerstrasse 5, CH-3012 Bern, Switzerland}

\begin{document} 
\maketitle

\vspace{-0.6cm}

\begin{abstract} \normalsize

We formulate lattice theories in which chiral symmetry is realized nonlinearly
on the fermion fields. In this framework the fermion mass term does not break 
chiral symmetry. This property allows us to use the Wilson term to remove the 
doubler fermions while maintaining exact chiral symmetry on the lattice. Our 
lattice formulation enables us to address non-perturbative questions in 
effective field theories of baryons interacting with pions and in models 
involving constituent quarks interacting with pions and gluons. We show that a 
system containing a non-zero density of static baryons interacting with pions 
can be studied on the lattice without encountering complex action problems. In 
our formulation one can also decide non-perturbatively if the chiral quark 
model of Georgi and Manohar provides an appropriate low-energy description of 
QCD. If so, one could understand why the non-relativistic quark model works.

\end{abstract}

\vspace{0.4cm}




\noindent
{\it PACS numbers}: 
%
%
11.30.Rd, 
12.38.Gc, 
12.38.Mh, 
12.39.Fe  
%
%
%
%
%

\maketitle
 
\newpage

\section{Introduction}

Realizing chiral symmetry on the lattice is a problem with a long
history. In 1975 Wilson decided to break chiral symmetry explicitly in
order to remove the unwanted doubler fermions \cite{Wil75}. As a
result, recovering chiral symmetry with Wilson fermions requires
fine-tuning as well as taking the continuum limit. In 1981 it was
proved by Nielsen and Ninomiya \cite{Nie81} and in a different way by
Friedan \cite{Fri82} that removing the doubler fermions while
maintaining locality is possible only if one explicitly breaks the
chiral symmetry of the continuum theory. In 1992, using a Wilson term
in 4+1 dimensions, Kaplan constructed domain wall fermions in order to
realize chiral symmetry on the lattice \cite{Kap92}. At the end of the
same year, Narayanan and Neuberger introduced the related overlap
formulation of chiral lattice fermions \cite{Nar93}.  Already in 1982
Ginsparg and Wilson \cite{Gin82} had discussed how to maintain a
remnant of chiral symmetry on the lattice. This paper was rediscovered
by Hasenfratz in 1997 in relation to the perfect action approach to
QCD \cite{Has98}.  In particular, perfect lattice fermions obey the by
now well-known Ginsparg-Wilson relation. It was immediately realized
by Neuberger that the same is true for overlap fermions \cite{Neu98}.
In 1998 L\"uscher showed that lattice fermion actions that obey the
Ginsparg-Wilson relation realize a new version of chiral symmetry that
is natural for theories on the lattice and coincides with the usual
one in the continuum limit \cite{Lue98}.  This development culminated
in the non-perturbative construction of chiral lattice gauge theories
\cite{Lue99}.

In this paper we consider another aspect of lattice chiral symmetry, namely
how it can be realized nonlinearly. Nonlinear realizations of symmetries are
well understood in the continuum in the context of low-energy effective
theories describing spontaneous symmetry breaking. Goldstone's theorem tells
us that the spontaneous breakdown of a continuous global symmetry $G$ to an
unbroken subgroup $H$ implies the presence of massless particles: the number
of Goldstone boson fields is given by the difference between the number of
generators of $G$ and $H$ \cite{Gol61}. The chiral symmetry group $G =
SU(N_f)_L \otimes SU(N_f)_R \otimes U(1)_B$ of QCD with $N_f$ massless quark
flavors is spontaneously broken to the vector subgroup $H = SU(N_f)_{L=R}
\otimes U(1)_B$.  Hence there are $N_f^2 - 1$ massless Goldstone bosons: the
three pions for $N_f = 2$ and, in addition, the four kaons, and the
$\eta$-meson for $N_f = 3$.  The theoretical framework for describing
Goldstone bosons with effective fields was developed by Weinberg \cite{Wei67}
and by Coleman, Wess, and Zumino \cite{Col69}, partly in collaboration with
Callan \cite{Cal69}. As a result, Goldstone bosons are described by fields in
the coset space $G/H$ and the group $G$ is nonlinearly realized on those
fields. In QCD the pions, the kaons, and the $\eta$-meson are accordingly
described by matrix valued fields in the coset space $G/H = SU(N_f)_L \otimes
SU(N_f)_R \otimes U(1)_B/ [SU(N_f)_{L=R} \otimes U(1)_B] = SU(N_f)$. Notably,
the pion effective theory developed by Weinberg \cite{Wei79} has a systematic
low-energy expansion in Gasser and Leutwyler's framework of chiral
perturbation theory \cite{Gas84}.

When matter fields are included in the effective theory of Goldstone bosons,
they also transform nonlinearly under chiral symmetry \cite{Geo84a}. In
particular, the global symmetry $G$ is hidden in a locally realized symmetry
$H$ with a corresponding ``gauge'' field that is constructed from the
Goldstone boson field. It must be pointed out that there is no true gauge
symmetry $H$, just the global symmetry $G$. Exploiting a nonlinearly realized
chiral symmetry, Gasser, Sainio, and Svarc incorporated a single nucleon in
chiral perturbation theory \cite{Gas88}. Systematic power counting schemes for
non-relativistic heavy baryon chiral perturbation theory were developed by
Jenkins and Manohar \cite{Jen91} and by Bernard, Kaiser, Kambor, and Meissner
\cite{Ber92}. A more economical scheme that maintains explicit Lorentz
invariance was introduced by Becher and Leutwyler \cite{Bec99}.

One may ask why one would want to put low-energy effective theories on the
lattice. Indeed, many questions that arise in the framework of low-energy
Lagrangians can be solved analytically using chiral perturbation theory.
Usually, such calculations are performed in the continuum using dimensional
regularization. It is certainly more tedious to work with the lattice
regularization in chiral perturbation theory calculations. Still, in order to
address conceptual problems, for example, related to power divergences, the
lattice has already been used in magnon \cite{Has93}, pion \cite{Smi99}, as
well as heavy baryon chiral perturbation theory \cite{Lew01}. In the context
of lattice QCD, effective field theories provide powerful tools to keep
systematic errors under control \cite{Kro03}. Moreover, chiral Lagrangians on
the lattice have been investigated in order to facilitate the lattice
determination of the Gasser-Leutwyler coefficients \cite{Myi94,Lev97}.
Although interesting, this is not the main motivation for the present work. We
formulate lattice theories with nonlinearly realized chiral symmetry in order
to address non-perturbative questions that arise in the context of low-energy
effective theories. It is perhaps not obvious that such problems even exist.
An example of a non-perturbative problem in the context of chiral Lagrangians
concerns the rotor spectrum of massless QCD in a finite box. However, this
problem reduces to one in quantum mechanics and has been solved analytically
by Leutwyler \cite{Leu87} without any need for a lattice. Few-nucleon systems
have been treated with effective field theory methods in nuclear physics. For
example, the two-nucleon system was studied by Kaplan, Savage, and Wise
\cite{Kap98} and the three-nucleon system was investigated by Bedaque, Hammer,
and van Kolck \cite{Bed98,Bed02}. Again, the solution of these problems
required non-perturbative techniques but not yet a lattice. Going to larger
nuclei and eventually to nuclear matter is a non-trivial task that requires to
solve more complicated non-perturbative problems in the effective theory
\cite{Mue00}. Here the lattice provides a natural framework in which such
questions can be addressed on a solid theoretical basis.

A central observation of this paper is that the Nielsen-Ninomiya theorem is
naturally avoided when chiral symmetry is nonlinearly realized on the lattice.
In fact, constituent quark masses do not break a nonlinearly realized 
chiral symmetry. Similarly, the Wilson term --- introduced in order to remove
the doubler fermions --- does no longer break chiral symmetry explicitly. 
Hence, exact chiral symmetry is naturally maintained using a simple Wilson 
action while the fermion doubling problem is still solved. This raises the 
question how anomalies manifest themselves in this context. Remarkably, in 
effective theories with nonlinearly realized chiral symmetry, both in the 
continuum and on the lattice, the fermion fields do not contribute to 
anomalies. Instead anomalies must be added by hand in the form of 
Wess-Zumino-Witten \cite{Wes71,Wit83} and Goldstone-Wilczek \cite{Gol81} terms.
Constructing these terms on the lattice while maintaining their topological 
features is a non-trivial problem which we do not address in this paper. In 
this respect, our
construction of lattice theories with nonlinearly realized chiral symmetry is 
still incomplete. Consequently, we limit ourselves to applications where 
anomalies play no major role. Such applications concern, for example, models of
static baryons interacting with pions at non-zero chemical potential and the 
chiral quark model of Georgi and Manohar \cite{Geo84b}. 

Dynamical fermion simulations in a lattice theory with nonlinearly realized 
chiral symmetry may be substantially easier than with the usual linear 
realization. Using the nonlinear realization, even in the chiral limit, the 
fermion mass is large due to spontaneous chiral symmetry breaking. Hence, the 
simulated dynamical fermions should behave algorithmically like heavy Wilson 
fermions while still maintaining manifest chiral symmetry. Using a linear 
realization of chiral symmetry on the quark fields, explicit auxiliary pion 
fields have already been explored in order to simplify QCD simulations 
\cite{Bro94,Kog98}. It has to be mentioned that, in general, lattice theories 
with nonlinearly realized chiral symmetry have a potential sign problem similar
to the one that arises for a single flavor of Wilson fermions. Fortunately, 
this sign problem is a lattice artifact which should disappear in the continuum
limit, in particular, for sufficiently heavy fermions. Recently, without 
directly addressing the potential sign problem, dynamical fermion algorithms 
have been developed for an odd number of flavors of Wilson fermions 
\cite{Bor95,Ale99,Tak01,Aok02}. We expect that similar methods are 
applicable to lattice fermions with nonlinearly realized chiral symmetry. 
Still, it should be pointed out that the feasibility of dynamical fermion 
simulations depends crucially on the severity of the sign problem, which should
be investigated in numerical simulations. We also need to mention that 
additional complex action problems arise if pseudo-vector fermion couplings or 
Wess-Zumino-Witten and Goldstone-Wilczek terms are included. In this paper we 
concentrate on the theoretical formulation of lattice theories with a 
nonlinearly realized chiral symmetry, while results of numerical simulations 
will be presented elsewhere.

Interestingly, the lattice construction presented here enables us to simulate 
massless pions interacting with static baryons even at non-zero baryon chemical
potential $\mu$ and finite temperature $T$, both for two and for three flavors.
In contrast to full QCD, in the static baryon model this is possible without 
encountering a complex action problem. Although questions involving high 
temperatures or high baryon densities are beyond the applicability of a 
systematic low-energy expansion, this model can still be used to answer 
questions concerning the universal features of the critical endpoint of the 
chiral phase transition line in the $(\mu,T)$-plane. Although one should not 
expect to obtain precise information on non-universal features of QCD, such as 
the location of the critical endpoint, one may learn at least 
semi-quantitatively how the location of the endpoint changes as a function of 
the strange quark mass. In addition, one will gain experience with numerical 
methods for locating the critical endpoint in a model that has the same global
symmetry properties as full QCD but is much simpler to simulate.

A nonlinear realization of chiral symmetry was also used by Georgi and 
Manohar in the construction of their chiral quark model \cite{Geo84b}. This
model contains massive constituent quarks interacting with pions and gluons. 
While gluons are flavor singlets, both constituent quarks and pions transform 
non-trivially under the nonlinearly realized chiral symmetry. If this model 
provides an appropriate low-energy description for QCD, one could understand 
why the non-relativistic quark model works. However, the confinement of 
constituent quarks is a non-perturbative issue that cannot be addressed 
quantitatively in the continuum formulation of the model. In this paper we 
present a lattice construction of the Georgi-Manohar model with nonlinearly
realized chiral symmetry. This non-perturbative framework will allow one to 
decide quantitatively if the Georgi-Manohar model can explain the success of 
the non-relativistic quark model.

The rest of this paper is organized as follows. In section 2 the nonlinear
realization of chiral symmetry is reviewed in the continuum, while section 3
contains the corresponding construction on the lattice. In both cases the 
relevant symmetries are discussed in detail. In particular, effective lattice 
theories for pions and baryons are considered and the issue of fermion doubling
is discussed. In section 4 we discuss how low-energy effective theories can be
treated non-perturbatively on the lattice and how continuum results can be
extracted. Section 5 explains how theories with static baryons can be 
simulated at non-zero chemical potential without encountering a complex action
problem. In section 6 we discuss the lattice version of the Georgi-Manohar 
model and propose numerical simulations in order to decide if it indeed 
provides a satisfactory explanation for the success of the non-relativistic 
quark model. Section 7 discusses the possible form of the phase diagram of the 
lattice chiral quark model and the relation of this model to Wilson's standard 
lattice QCD. Finally, section 8 contains our conclusions.

\section{Continuum Formulation of Nonlinearly \\ Realized Chiral Symmetry}

In this section we review how theories with nonlinearly realized chiral 
symmetry are formulated in the continuum. The fields used in this construction
are introduced with emphasis on their behavior under chiral symmetry 
transformations, charge conjugation, and parity. Simple examples of low-energy
effective theories with nonlinearly realized chiral symmetry are given and
manifestations of anomalies are being discussed.

Consider the chiral symmetry group $G = SU(N_f)_L \otimes SU(N_f)_R \otimes 
U(1)_B$ of QCD involving $N_f$ quark flavors which breaks spontaneously to the 
subgroup $H = SU(N_f)_{L=R} \otimes U(1)_B$. Following \cite{Col69,Cal69} the 
Goldstone boson field is then given by $U(x) = \exp[2 i \pi^a(x) T^a/F_\pi] \in
G/H = SU(N_f)$. Here $\mbox{Tr}(T^a T^b) = 2 \delta_{ab}$ and $F_\pi$ is the 
pion decay constant. Under global chiral rotations $L \otimes R \in SU(N_f)_L 
\otimes SU(N_f)_R$, this field transforms as
\begin{equation}
U(x)' = L U(x) R^\dagger.
\end{equation}
From the Goldstone boson field one derives a field $u(x) \in SU(N_f)$ as
\begin{equation}
u(x) = U(x)^{1/2}.
\end{equation}
In order to avoid ambiguities in taking the square-root, we proceed as follows.
First, the field $U(x)$ is diagonalized by a unitary transformation $W(x)$, 
i.e.
\begin{equation}
U(x) = W(x) D(x) W(x)^\dagger,
\end{equation}
where $D(x)$ is a diagonal matrix of complex phases
\begin{equation}
D(x) = \mbox{diag}\{\exp[i \varphi_1(x)],\exp[i \varphi_2(x)],...,
\exp[i \varphi_{N_f}(x)]\}.
\end{equation}
At this point, the phases $\varphi_i(x)$ are defined only modulo $2 \pi$. In
order to fix this ambiguity, we demand that
\begin{equation}
\sum_{i=1}^{N_f} \varphi_i(x) = 0, \ |\varphi_i(x) - \varphi_j(x)| \leq 2 \pi.
\end{equation}
This uniquely fixes the phases of the eigenvalues and defines the matrix $D(x)$
up to an irrelevant permutation of its diagonal elements. We then define
\begin{equation}
D(x)^{1/2} = \mbox{diag}\{\exp[i \varphi_1(x)/2],\exp[i \varphi_2(x)/2],...,
\exp[i \varphi_{N_f}(x)/2]\},
\end{equation}
and we obtain
\begin{equation}
u(x) = W(x) D(x)^{1/2} W(x)^\dagger.
\end{equation}
When the eigenvalues of $U(x)$ are non-degenerate, the matrix $u(x)$ is
uniquely defined in this way and is located in the middle of the shortest
geodesic connecting $U(x)$ with the unit-matrix $\1$ in the group manifold of
$SU(N_f)$. On the other hand, if two eigenvalues are degenerate, the
corresponding eigenvectors are not uniquely defined. If, in addition, the
corresponding phases $\varphi_i$ and $\varphi_j$ differ by $2 \pi$, this
ambiguity gives rise to a coordinate singularity in the definition of $u(x)$.
In the $N_f = 2$ case, the singularities arise for $U(x) = - \1$, which
corresponds to a single point in the $SU(2)$ group manifold. For general $N_f$,
the submanifold of $SU(N_f)$ matrices with two degenerate eigenvalues (and with
$|\varphi_i - \varphi_j| = 2 \pi$) has dimension $N_f^2 - 4$. Hence, the
codimension of the singular submanifold is $N_f^2 - 1 - (N_f^2 - 4) = 3$ for
all $N_f$. Consequently, in a 4-dimensional space-time the singularities in
$u(x)$ are located on 1-dimensional lines. If one chooses the vacuum as
$U(x) = \1$, the singularities represent the world-lines of Skyrmion centers.
However, it should be noted that the location of the singularities varies under
those chiral rotations that do not leave the vacuum invariant. Still, the
effective actions that we will construct with $u(x)$ are completely chirally
invariant, even in the presence of Skyrmions or other large-field
configurations.

Under chiral rotations the field $u(x)$ transforms as
\begin{equation}
u(x)' = L u(x) V(x)^\dagger = V(x) u(x) R^\dagger.
\end{equation}
The matrix $V(x)$ depends on $L$ and $R$ as well as on the field $U(x)$ and can
be written as
\begin{equation}
V(x) = R[R^\dagger L U(x)]^{1/2} [U(x)^{1/2}]^\dagger =
L[L^\dagger R U(x)^\dagger]^{1/2} U(x)^{1/2}.
\end{equation}
For transformations in the unbroken vector subgroup $SU(N_f)_{L=R}$, $L = R$,  
the field $V(x)$ reduces to the global flavor transformation $V(x) = L = R$. 
The local transformation $V(x)$ is a nonlinear representation of chiral 
symmetry. For two subsequent chiral transformations with $L_2 L_1 = L$ and 
$R_2 R_1 = R$, one has
\begin{equation}
\label{V1}
V_1(x) = R_1[R_1^\dagger L_1 U(x)]^{1/2} [U(x)^{1/2}]^\dagger, \
V_2(x) = R_2[R_2^\dagger L_2 U(x)']^{1/2} [{U(x)'}^{1/2}]^\dagger,
\end{equation}
with $U(x)' = L_1 U(x) R_1^\dagger$, so that indeed
\begin{eqnarray}
\label{V2}
V_2(x) V_1(x)&=&R_2[R_2^\dagger L_2 L_1 U(x) R_1^\dagger]^{1/2} 
R_1 [U(x)^{1/2}]^\dagger \nonumber \\
&=&R_2 R_1 [R_1^\dagger R_2^\dagger L_2 L_1 U(x)]^{1/2}
[U(x)^{1/2}]^\dagger \nonumber \\
&=&R[R^\dagger L U(x)]^{1/2} [U(x)^{1/2}]^\dagger = V(x).
\end{eqnarray}

Now consider a Dirac spinor field $\Psi(x)$ and $\overline \Psi(x)$ that
transforms under the fundamental representation of the flavor group. This field
may represent a nucleon (for $N_f = 2$) or a constituent quark (for any value 
of $N_f$). Global chiral rotations $L \otimes R \in SU(N_f)_L \otimes 
SU(N_f)_R$ can be realized nonlinearly on this field using the transformations
\begin{equation}
\Psi(x)' = V(x) \Psi(x), \ \overline \Psi(x)' = \overline \Psi(x) V(x)^\dagger.
\end{equation}
This has the form of an $SU(N_f)$ gauge transformation despite the fact that it
represents just a global $SU(N_f)_L \otimes SU(N_f)_R$ symmetry. 

In order to construct a chirally invariant (i.e. $SU(N_f)$ ``gauge'' invariant)
Lagrangian, one needs an $SU(N_f)$ flavor ``gauge'' field. Indeed, the 
anti-Hermitean composite field
\begin{equation}
\label{v}
v_\mu(x) = \frac{1}{2}[u(x)^\dagger \partial_\mu u(x) + 
u(x) \partial_\mu u(x)^\dagger]
\end{equation}
transforms as a ``gauge'' field
\begin{eqnarray}
v_\mu(x)'&=&\frac{1}{2}
\{V(x) u(x)^\dagger L^\dagger \partial_\mu [L u(x) V(x)^\dagger] +
V(x) u(x) R^\dagger \partial_\mu [R u(x)^\dagger V(x)^\dagger]\} \nonumber \\
&=&V(x) [v_\mu(x) + \partial_\mu] V(x)^\dagger.
\end{eqnarray}
A Hermitean composite field is given by
\begin{equation}
\label{a}
a_\mu(x) = \frac{i}{2}[u(x)^\dagger \partial_\mu u(x) - 
u(x) \partial_\mu u(x)^\dagger],
\end{equation}
which transforms as
\begin{eqnarray}
a_\mu(x)'&=&\frac{1}{2}
\{V(x) u(x)^\dagger L^\dagger \partial_\mu [L u(x) V(x)^\dagger] -
V(x) u(x) R^\dagger \partial_\mu [R u(x)^\dagger V(x)^\dagger]\} \nonumber \\ 
&=&V(x) a_\mu(x) V(x)^\dagger.
\end{eqnarray}


It is interesting to investigate the transformation properties of the fields
$\Psi(x)$, $\overline \Psi(x)$, $U(x)$, $u(x)$, $v_\mu(x)$, and $a_\mu(x)$ 
under charge conjugation and parity. Under charge conjugation, the fermion 
field transforms in the usual way
\begin{equation}
\label{cg}
^C\Psi(x) = C \overline \Psi(x)^T, \ ^C\overline \Psi(x) = - \Psi(x)^T C^{-1},
\end{equation}
with $C^{-1} \gamma_\mu C = - \gamma_\mu^T$. The Goldstone boson field $U(x)$ 
transforms as
\begin{equation}
^CU(x) = U(x)^T.
\end{equation}
Consequently, the field $u(x)$ also follows the same transformation law
\begin{equation}
^Cu(x) = [^CU(x)]^{1/2} = [U(x)^T]^{1/2} = [U(x)^{1/2}]^T = u(x)^T.
\end{equation}
It is easy to show that the composite fields $v_\mu(x)$ and $a_\mu(x)$ 
transform as
\begin{equation}
^Cv_\mu(x) = v_\mu(x)^*, \ ^Ca_\mu(x) = a_\mu(x)^*.
\end{equation}

Under parity, the fermion field behaves again in the usual way, i.e.
\begin{equation}
^P\Psi(\vec x,t) = \gamma_4 \Psi(- \vec x,t), \ 
^P\overline \Psi(\vec x,t) = \overline \Psi(- \vec x,t) \gamma_4,
\end{equation}
where 4 denotes the Euclidean time direction. The Goldstone boson field $U(x)$ 
transforms as
\begin{equation}
^PU(\vec x,t) = U(- \vec x,t)^\dagger
\end{equation}
so that
\begin{equation}
^Pu(\vec x,t) = [^PU(\vec x,t)]^{1/2} = [U(- \vec x,t)^\dagger]^{1/2} = 
[U(- \vec x,t)^{1/2}]^\dagger = u(- \vec x,t)^\dagger.
\end{equation}
Similarly, one finds
\begin{eqnarray}
&&^Pv_i(\vec x,t) = - v_i(- \vec x,t), \ ^Pv_4(\vec x,t) = v_4(- \vec x,t), 
\nonumber \\
&&^Pa_i(\vec x,t) = a_i(- \vec x,t), \ ^Pa_4(\vec x,t) = - a_4(- \vec x,t), 
\end{eqnarray}
where $i \in \{1,2,3\}$ denotes a spatial direction. Hence, $v_\mu(x)$ is a 
vector and $a_\mu(x)$ is an axial vector.

At this point we are prepared to write down the leading terms in the Euclidean
action of a low-energy effective theory for nucleons (with $N_f = 2$)
\begin{eqnarray}
\label{action2}
S[\overline \Psi,\Psi,U]&=&\int d^4x \
\Big\{M \overline \Psi \Psi + 
\overline \Psi \gamma_\mu (\partial_\mu + v_\mu) \Psi +
i g_A \overline \Psi \gamma_\mu \gamma_5 a_\mu \Psi \nonumber \\
&+&\frac{F_\pi^2}{4} \mbox{Tr}[\partial_\mu U^\dagger \partial_\mu U] -
\frac{\langle \overline \psi \psi \rangle}{2} \mbox{Tr}[{\cal M} U^\dagger +
{\cal M}^\dagger U]\Big\}.
\end{eqnarray}
Here $M$ is the nucleon mass generated by spontaneous chiral symmetry breaking,
$g_A$ is the coupling to the isovector axial current, and 
$\langle \overline \psi \psi \rangle$ and ${\cal M} = \mbox{diag}(m_u,m_d)$ are
respectively the chiral condensate per flavor and the mass matrix of the 
current quarks. When $\Psi(x)$ represents a constituent quark, $M$ would be the
constituent quark mass. In that case, gluons must also be included in order to 
confine the constituent quarks. This is what happens in the chiral quark model 
of Georgi and Manohar \cite{Geo84b} (see section 6).

It is remarkable that --- due to the nonlinear realization of chiral symmetry 
--- the fermion mass term is chirally invariant. This makes sense because the 
mass $M$ arises from the spontaneous breakdown of chiral symmetry even in the 
chiral limit, i.e. when the current quarks are massless. In fact, the term that
contains the matrix ${\cal M}$ of current quark masses is the only source of 
explicit chiral symmetry breaking in the leading terms of the effective theory.
It is interesting to note that the composite field $v_\mu(x)$ is coupled to the
fermions as a flavor ``gauge'' field that enters the 
covariant derivative $D_\mu = \partial_\mu + v_\mu$. Like $v_\mu$, the axial 
vector field $a_\mu$ also transforms in the adjoint representation of the 
flavor group. However, it is not a ``gauge'' field and it couples to the 
fermions with strength $g_A$. It must be noted that there is, of course, no 
true $SU(N_f)$ gauge symmetry, just an $SU(N_f)_L \otimes SU(N_f)_R \otimes 
U(1)_B$ global symmetry. In particular, the ``gauge'' field $v_\mu(x)$ is not a
fundamental field and there is, for instance, no $SU(N_f)$ Gauss law.

Until now we have considered a fermion field $\Psi(x)$ transforming in the 
fundamental representation of $SU(N_f)$. In QCD with three massless flavors 
($N_f = 3$) the baryons transform as a flavor octet. The corresponding Dirac 
spinor field, denoted by $\chi(x)$ and $\overline \chi(x)$, is traceless and 
transforms in the adjoint representation
\begin{equation}
\chi(x)' = V(x) \chi(x) V(x)^\dagger.
\end{equation}
The covariant derivative is then given by
\begin{equation}
D_\mu \chi(x) = \partial_\mu \chi(x) + [v_\mu(x),\chi(x)],
\end{equation}
which transforms as $D_\mu \chi(x)' = V(x) D_\mu \chi(x) V(x)^\dagger$ under 
the nonlinearly realized chiral symmetry. In this case, the Euclidean action 
of the effective theory for pions, kaons, $\eta$-mesons, and baryons takes the 
form
\begin{eqnarray}
\label{Nf3action}
S[\overline \chi,\chi,U]&=&\int d^4x \
\Big\{\mbox{Tr}[M \overline \chi \chi + \overline \chi \gamma_\mu D_\mu \chi +
i D \overline \chi \gamma_\mu \gamma_5 \{a_\mu,\chi\} +
i F \overline \chi \gamma_\mu \gamma_5 [a_\mu,\chi]] \nonumber \\
&+&\frac{F_\pi^2}{4} \mbox{Tr}[\partial_\mu U^\dagger \partial_\mu U] -
\frac{\langle \overline \psi \psi \rangle}{2} \mbox{Tr}[{\cal M} U^\dagger +
{\cal M}^\dagger U]\Big\},
\end{eqnarray}
where the matrix ${\cal M} = \mbox{diag}(m_u,m_d,m_s)$ now also includes the 
strange quark mass. The low-energy parameters $D$ and $F$ are analogous to the 
coupling $g_A$ of the two flavor case.

It is interesting to discuss how anomalies arise in effective theories with
nonlinearly realized chiral symmetry. Let us begin with a discussion of the 
axial $U(1)_A$ symmetry of QCD. When the number of colors $N_c$ is finite, this
symmetry of the classical action is anomalously broken by quantum effects due 
to instantons \cite{tHo76}. As a result, the $\eta'$-meson is not a Goldstone
boson. Only in the large $N_c$ limit the $\eta'$-meson becomes a 
pseudo-Goldstone boson \cite{Ven79,Wit79}. For $N_c = 3$, the $U(1)_A$ symmetry
is so strongly broken that it cannot even be considered an approximate symmetry
of QCD. In particular, the $\eta'$-meson is so heavy that it is not included 
explicitly in the low-energy effective theory. Indeed, the effective theories 
that we have written down do not have a $U(1)_A$ symmetry at all. This follows 
because the Goldstone boson field lives in $SU(N_f)$ and not in $U(N_f)$. 

Let us now move on to those anomalies that arise when one gauges the
electroweak interactions. In effective theories with nonlinearly realized 
chiral symmetries anomalies manifest themselves \cite{Geo84a} in the form of 
Wess-Zumino-Witten \cite{Wes71,Wit83} and Goldstone-Wilczek terms \cite{Gol81}.
For example, in the $N_f = 2$ case, Witten's global $SU(2)_L$ anomaly 
\cite{Wit82} manifests itself in a topological factor 
$\mbox{Sign}[U]^{N_c} \in \Pi_4[SU(2)] = \{1,-1\}$ that must be included in the
path integral \cite{DHo84}. This ensures that, for odd $N_c$, the Skyrme 
soliton \cite{Sky61} is quantized as a fermion representing the physical 
nucleon in the pure pion theory. The $SU(2)_L$ gauge invariant version of the 
Skyrme baryon number current is the Goldstone-Wilczek current \cite{Gol81}. In 
the presence of electroweak instantons, this current is no longer conserved and 
reproduces the 't~Hooft anomaly in the baryon number \cite{tHo76}.

In the $N_f \geq 3$ case, the Wess-Zumino-Witten term replaces the topological
factor $\mbox{Sign}[U]$ in the Goldstone boson theory. This is because then 
$\Pi_4[SU(N_f)] = \{0\}$ and instead $\Pi_5[SU(N_f)] = \Z$. The 
Wess-Zumino-Witten action has the form
\begin{equation}
S_{WZW}[U] = \frac{1}{240 \pi^2 i} \int_{H^5} d^5x \ 
\varepsilon_{\mu\nu\rho\sigma\lambda} \mbox{Tr}[(U^\dagger \p_\mu U)
(U^\dagger \p_\nu U)(U^\dagger \p_\rho U)(U^\dagger \p_\sigma U)
(U^\dagger \p_\lambda U)],
\end{equation}
where $H^5$ is a 5-dimensional hemi-sphere whose boundary $\p H^5 = S^4$ is 
(compactified) space-time. The physical Goldstone boson field 
$U(x) \in SU(N_f)$ is extended to a field $U(x,x_5)$ that depends on the 
unphysical fifth coordinate $x_5 \in [0,1]$ such that $U(x,0) = \1$ and 
$U(x,1) = U(x)$. Of course, the 4-dimensional physics should be independent 
of how the field $U(x,x_5)$ is deformed into the bulk of the fifth dimension,
and should only depend on $U(x)$. This is possible because the interpolation
ambiguity of $S_{WZW}[U]$ is given by $2 \pi$ times a winding number in 
$\Pi_5[SU(N_f)] = \Z$. Hence, when the prefactor of the Wess-Zumino-Witten 
term is quantized in integer units, the 4-dimensional physics is independent of
how one interpolates the Goldstone boson field into the unphysical fifth 
dimension. Remarkably, in the low-energy theory for QCD, the quantized 
prefactor is the number of colors $N_c$ so that
\begin{equation}
Z = \int {\cal D}U \exp(- S[U]) \exp(i N_c S_{WZW}[U]).
\end{equation}
For $N_f = 2$, one can show that 
$\exp(i N_c S_{WZW}[U]) = \mbox{Sign}[U]^{N_c}$. The Wess-Zumino-Witten term
contains the vertex that describes the anomalous decay of the neutral pion into
two photons. For general $N_c$, this process also receives contributions from a
Goldstone-Wilczek term \cite{Bae01}.

In contrast to the $N_f = 2$ case, for $N_f \geq 3$, the Wess-Zumino-Witten 
term also breaks an unwanted intrinsic parity symmetry $P_0$ that is present in
the Goldstone boson action $S[U]$ but not in QCD \cite{Wit83}. The full parity 
$P$ acts on the pseudo-scalar Goldstone bosons $\pi^a(\vec x,t)$ by spatial 
inversion accompanied by a sign-change, i.e.\ 
$^P\pi^a(\vec x,t) = - \pi^a(- \vec x,t)$, such that
\begin{equation}
^PU(\vec x,t) = U(- \vec x,t)^\dagger.
\end{equation}
The intrinsic parity $P_0$, on the other hand, leaves out the spatial inversion
and takes the form
\begin{equation}
^{P_0}U(\vec x,t) = U(\vec x,t)^\dagger.
\end{equation}
If $P_0$ were a symmetry of QCD, the number of Goldstone bosons would be
conserved modulo two. While this is indeed the case for $N_f = 2$ (where $P_0$ 
is nothing but $G$-parity \cite{Lee56}), for $N_f \geq 3$ intrinsic parity is 
not a symmetry of QCD. For example, the $\phi$-meson can decay both into two 
kaons or into three pions. However, the Goldstone boson action $S[U]$ is indeed
invariant under $P_0$,
\begin{equation}
S[^{P_0}U] = S[U^\dagger] = S[U].
\end{equation}
Hence it has more symmetry than the underlying QCD action. The 
Wess-Zumino-Witten action, on the other hand, is odd under $P_0$,
\begin{equation}
S_{WZW}[^{P_0}U] = S_{WZW}[U^\dagger] = - S_{WZW}[U],
\end{equation}
and thus reduces the symmetry of the effective theory to the one of QCD.

The next step is to ask how anomalies are represented when explicit baryon 
fields are present in the low-energy effective theory. In the presence of 
electroweak $SU(2)_L \otimes U(1)_Y$ gauge fields 
$W_\mu(x) = i g W_\mu^a(x) T^a$ and $B_\mu(x) = i g' B_\mu^3(x) T^3$ one
obtains
\begin{eqnarray}
v_\mu(x) = \frac{1}{2}\{u(x)^\dagger[\p_\mu + W_\mu(x)]u(x) +
u(x)[\p_\mu + B_\mu(x)]u(x)^\dagger\}, \nonumber \\
a_\mu(x) = \frac{i}{2}\{u(x)^\dagger[\p_\mu + W_\mu(x)]u(x) -
u(x)[\p_\mu + B_\mu(x)]u(x)^\dagger\},
\end{eqnarray}
which still transform as
\begin{equation}
v_\mu(x)' = V(x)[v_\mu(x) + \p_\mu]V(x)^\dagger, \ 
a_\mu(x)' = V(x) a_\mu(x) V(x)^\dagger.
\end{equation}
Hence, the dynamical gauge fields $W_\mu(x)$ and $B_\mu(x)$ are incorporated in
the ``gauge'' field $v_\mu(x)$ constructed from the pion field $U(x)$. By 
varying the resulting action with respect to $W_\mu(x)$ and $B_\mu(x)$, one 
obtains the left- and right-handed fermion currents
\begin{eqnarray}
&&j_{\mu L}^a(x) = \overline \Psi(x) u(x)^\dagger \gamma_\mu
\frac{1 - g_A \gamma_5}{2} T^a u(x) \Psi(x), \nonumber \\
&&j_{\mu R}^a(x) = \overline \Psi(x) u(x) \gamma_\mu
\frac{1 + g_A \gamma_5}{2} T^a u(x)^\dagger \Psi(x).
\end{eqnarray}
Since fermions with a nonlinearly realized chiral symmetry do not contribute to
anomalies \cite{Geo84a,Man84}, even in the presence of fermion fields, the 
anomalies are still contained in Goldstone-Wilczek and Wess-Zumino-Witten 
terms. 

With explicit baryon fields being present in addition to the Skyrme solitons in
the pion field, one may wonder if baryons are doubly counted in this approach. 
First of all, in the power counting scheme of chiral perturbation theory one 
cannot address questions about Skyrmions because all contributions from higher
order terms are then equally important. In a non-perturbative approach to the 
effective theory, one can ask if, for example, in the sector with one baryon, 
the perturbative vacuum configuration of the pion field becomes unstable 
against Skyrmion formation. If so, the chiral power counting scheme is not 
applicable and the systematic low-energy approach fails. Hence, in the 
low-energy effective theory, one is restricted to the trivial topological 
sector and the issue of double counting does not arise. Since Skyrmions have no
place in the systematic low-energy expansion, one cannot use the 
Goldstone-Wilczek current to describe baryon number violating processes in the 
framework of chiral perturbation theory. This should not be too surprising. 
Baryon decay --- as rare as it may be --- is a violent event that releases a 
lot of energy and can hence not be understood within a low-energy effective 
theory. 

\section{Lattice Formulation of Nonlinearly Realized \\ Chiral Symmetry}

Let us now construct theories with a nonlinearly realized chiral symmetry on 
the lattice. In analogy to the previous section, we set up the lattice fields 
needed for this construction such that the correct symmetry properties are 
respected. With these fields, we formulate lattice actions of low-energy 
effective theories in which chiral symmetry is nonlinearly realized. In 
particular, we discuss how fermion doublers can be removed using a Wilson term 
without explicitly breaking chiral symmetry and how the Nielsen-Ninomiya 
theorem is avoided. The issue of anomalies is also addressed in the lattice 
formulation.

The Goldstone boson field $U_x \in SU(N_f)$ lives on the sites $x$ of 
a 4-dimensional hyper-cubic lattice and it transforms under global chiral
rotations as
\begin{equation}
U_x' = L U_x R^\dagger.
\end{equation}
As in the continuum, the field $u_x \in SU(N_f)$ is constructed as 
$u_x = U_x^{1/2}$ and it transforms as
\begin{equation}
u_x' = L u_x V_x^\dagger = V_x u_x R^\dagger.
\end{equation}
Again, the field $V_x$ depends on the global chiral transformations $L$ and $R$
as well as on the field $U_x$ and is given by
\begin{equation}
V_x = R(R^\dagger L U_x)^{1/2} (U_x^{1/2})^\dagger =
L(L^\dagger R U_x^\dagger)^{1/2} U_x^{1/2}.
\end{equation}
The continuum results for two subsequent chiral transformations given in 
eqs.~(\ref{V1}) and (\ref{V2}) remain completely unchanged on the lattice. 
Also the discussion of the coordinate singularities still applies. However, 
it should be noted that the singularities generically fall between lattice 
points. Still, using an appropriate interpolation, it is possible to detect 
these topological objects even on the lattice.

Next we consider the fermion field $\Psi_x$ and $\overline \Psi_x$ which again 
lives on the lattice sites $x$. As in the continuum, global chiral rotations 
are nonlinearly realized on this field such that
\begin{equation}
\Psi_x' = V_x \Psi_x, \ \overline \Psi_x' = \overline \Psi_x V_x^\dagger.
\end{equation}
This field transforms in the fundamental representation of $SU(N_f)$ and can
hence represent a nucleon (for $N_f = 2$) or a constituent quark (for any value
of $N_f$). For $N_f = 3$, the lattice baryon field $\chi_x$ and 
$\overline \chi_x$ also lives on the lattice sites but transforms in the 
adjoint representation
\begin{equation}
\chi_x' = V_x \chi_x V_x^\dagger, \ 
\overline \chi_x' = V_x \overline \chi_x V_x^\dagger.
\end{equation}

We now proceed to the construction of $V_{x,\mu} \in SU(N_f)$ --- the lattice 
analog of the continuum flavor ``gauge'' field $v_\mu(x)$. It should be noted
that $V_{x,\mu}$ is a flavor parallel transporter along a lattice link in the 
group $SU(N_f)$, while the continuum flavor ``gauge'' field is in the algebra
$su(N_f)$. In analogy to the continuum expression (\ref{v}) for $v_\mu(x)$, we 
construct the lattice object
\begin{equation}
\tilde V_{x,\mu} = \frac{1}{2}[u_x^\dagger u_{x+\hat\mu} + 
u_x u_{x+\hat\mu}^\dagger].
\end{equation}
Here $\hat\mu$ is the unit-vector in the $\mu$-direction. By construction, 
$\tilde V_{x,\mu}$ transforms as a parallel transporter under the nonlinearly 
realized chiral symmetry, i.e.
\begin{equation}
\tilde V_{x,\mu}' = V_x \tilde V_{x,\mu} V_{x+\hat\mu}^\dagger.
\end{equation}
In the classical continuum limit, $\tilde V_{x,\mu} = 
\exp[v_\mu(x+\hat\mu/2)]$. However, at finite lattice spacing,
$\tilde V_{x,\mu}$ is in general not an element of $SU(N_f)$, just a complex 
$N_f \times N_f$ matrix in the group $GL(N_f)$. Hence, it is natural to project
a group-valued parallel transporter $V_{x,\mu} \in SU(N_f)$ out of 
$\tilde V_{x,\mu}$ by performing a $GL(N_f)/SU(N_f)$ coset decomposition
\begin{equation}
\tilde V_{x,\mu} = H_{x,\mu} \exp(i \varphi_{x,\mu}/N_f) V_{x,\mu}.
\end{equation}
Here
\begin{equation}
H_{x,\mu} = (\tilde V_{x,\mu} \tilde V_{x,\mu}^\dagger)^{1/2}
\end{equation}
is a positive Hermitean matrix that transforms as
\begin{equation}
H_{x,\mu}' = V_x H_{x,\mu} V_x^\dagger,
\end{equation}
and $\exp(i \varphi_{x,\mu})$ (with $\varphi_{x,\mu} \in \ ]\!-\pi,\pi]$) is 
the phase of the determinant of $\tilde V_{x,\mu}$. Then, by construction,
\begin{equation}
V_{x,\mu} = \exp(- i \varphi_{x,\mu}/N_f) H_{x,\mu}^{-1} 
\tilde V_{x,\mu},
\end{equation}
is indeed in $SU(N_f)$ and transforms as
\begin{equation}
\label{Vtrans}
V_{x,\mu}' = V_x V_{x,\mu} V_{x+\hat\mu}^\dagger.
\end{equation}

We also want to construct a lattice version of the continuum field $a_\mu(x)$
defined in (\ref{a}). For this purpose, we first consider
\begin{equation}
\tilde A_{x,\mu} = \frac{i}{2}[u_x^\dagger u_{x+\hat\mu} - 
u_x u_{x+\hat\mu}^\dagger].
\end{equation}
While in the continuum $a_\mu(x)' = V(x) a_\mu(x) V(x)^\dagger$, the lattice 
field $\tilde A_{x,\mu}$ transforms as
\begin{equation}
\tilde A_{x,\mu}' = V_x \tilde A_{x,\mu} V_{x+\hat\mu}^\dagger.
\end{equation}
Also, in contrast to the continuum field $a_\mu(x)$, the lattice field 
$\tilde A_{x,\mu}$ is in general neither traceless nor Hermitean. To cure these
problems, it is natural to introduce the field
\begin{equation}
\label{Aleft}
A^L_{x,\mu} = \frac{1}{2}[\tilde A_{x,\mu} V_{x,\mu}^\dagger +
V_{x,\mu} \tilde A_{x,\mu}^\dagger] - \frac{1}{2N_f} \mbox{Tr}
[\tilde A_{x,\mu} V_{x,\mu}^\dagger + V_{x,\mu} \tilde A_{x,\mu}^\dagger] \1,
\end{equation}
which, by construction, is indeed traceless and Hermitean and which transforms
as 
\begin{equation}
{A^L_{x,\mu}}' = V_x A^L_{x,\mu} V_x^\dagger,
\end{equation}
with the matrix $V_x$ located at the site $x$ on the left end of the link
$(x,\mu)$. Similarly, we define the object
\begin{equation}
\label{Aright}
A^R_{x,\mu} = \frac{1}{2}[V_{x,\mu}^\dagger \tilde A_{x,\mu} +
\tilde A_{x,\mu}^\dagger V_{x,\mu}] - \frac{1}{2N_f} \mbox{Tr}
[V_{x,\mu}^\dagger \tilde A_{x,\mu} + \tilde A_{x,\mu}^\dagger V_{x,\mu}] \1,
\end{equation}
which transforms as 
\begin{equation}
{A^R_{x,\mu}}' = V_{x+\hat\mu} A^R_{x,\mu} V_{x+\hat\mu}^\dagger,
\end{equation}
with the matrix $V_{x+\hat\mu}$ located at the site $x+\hat\mu$ on the right 
end of the link $(x,\mu)$. It should be noted that $A^R_{x,\mu}$ and
$A^L_{x,\mu}$ are not independent, but are related by parallel transport, i.e.
\begin{equation}
A^R_{x,\mu} = V_{x,\mu}^\dagger A^L_{x,\mu} V_{x,\mu}.
\end{equation}

Let us now consider the symmetry properties of the lattice fields under charge
conjugation and parity. The site fields $\Psi_x$, $\overline \Psi_x$, $\chi_x$,
$\overline \chi_x$, $U_x$, and $u_x$ transform exactly as their continuum 
analogs and will not be discussed again. Instead, we concentrate on the fields 
$V_{x,\mu}$, $A^L_{x,\mu}$, and $A^R_{x,\mu}$. Under charge conjugation, they 
transform as
\begin{equation}
^CV_{x,\mu} = V_{x,\mu}^*, \ ^CA^L_{x,\mu} = A^{L*}_{x,\mu}, \
^CA^R_{x,\mu} = A^{R*}_{x,\mu}.
\end{equation}
Similarly, under parity transformations, one finds
\begin{eqnarray}
&&^PV_{\vec x,t,i} = V_{- \vec x - \hat i,t,i}^\dagger \ , \ 
^PV_{\vec x,t,4} = V_{- \vec x,t,4} \ , \nonumber \\
&&^PA^L_{\vec x,t,i} = A^R_{- \vec x - \hat i,t,i} \ , \ 
^PA^L_{\vec x,t,4} = - A^L_{- \vec x,t,4} \ , \nonumber \\  
&&^PA^R_{\vec x,t,i} = A^L_{- \vec x,t,i} \ , \ 
^PA^R_{\vec x,t,4} = - A^R_{- \vec x,t,4} \ ,  
\end{eqnarray}
where $i \in \{1,2,3\}$ denotes a spatial direction and 4 again denotes the 
Euclidean time direction. Hence, as in the continuum, $V_{x,\mu}$ is a vector 
and $A^L_{x,\mu}$ and $A^R_{x,\mu}$ define an axial vector.

We now use the fields introduced above to construct lattice actions for baryon 
effective theories. We start out with the $N_f = 2$ case. In close analogy to 
the continuum action (\ref{action2}), we define the lattice action as
\begin{eqnarray}
\label{latact2}
S[\overline \Psi,\Psi,U]&=&M \sum_x \overline \Psi_x \Psi_x + 
\frac{1}{2} \sum_{x,\mu} 
(\overline \Psi_x \gamma_\mu V_{x,\mu} \Psi_{x+\hat\mu} -
\overline \Psi_{x+\hat\mu} \gamma_\mu V_{x,\mu}^\dagger \Psi_x)
\nonumber \\
&+&\frac{i g_A}{2} \sum_{x,\mu}  
(\overline \Psi_x \gamma_\mu \gamma_5 A^L_{x,\mu} \Psi_{x} +
\overline \Psi_{x+\hat\mu} \gamma_\mu \gamma_5 A^R_{x,\mu} \Psi_{x+\hat\mu})
\nonumber \\
&+&\frac{r}{2} \sum_{x,\mu} (2 \overline \Psi_x \Psi_x -
\overline \Psi_x V_{x,\mu} \Psi_{x+\hat\mu} -
\overline \Psi_{x+\hat\mu} V_{x,\mu}^\dagger \Psi_x)
\nonumber \\
&-&\frac{F_\pi^2}{4} \sum_{x,\mu} \mbox{Tr}[U_x^\dagger U_{x+\hat\mu} +
U_{x+\hat\mu}^\dagger U_x] - \frac{\langle \overline \psi \psi \rangle}{2}
\sum_x \mbox{Tr}[{\cal M} U_x^\dagger + {\cal M}^\dagger U_x]. \nonumber \\ \
\end{eqnarray}
Similarly, for $N_f = 3$, (where the baryon field transforms in the adjoint
representation) the lattice action corresponding to (\ref{Nf3action}) is given 
by
\begin{eqnarray}
\label{latact3}
S[\overline \chi,\chi,U]&=&M \sum_x \mbox{Tr}\overline \chi_x \chi_x + 
\frac{1}{2} \sum_{x,\mu} \mbox{Tr}
(\overline \chi_x \gamma_\mu V_{x,\mu} \chi_{x+\hat\mu} V_{x,\mu}^\dagger -
\overline \chi_{x+\hat\mu} \gamma_\mu V_{x,\mu}^\dagger \chi_x V_{x,\mu})
\nonumber \\
&+&\frac{i D}{2} \sum_{x,\mu} \mbox{Tr}
(\overline \chi_x \gamma_\mu \gamma_5 \{A^L_{x,\mu},\chi_{x}\} +
\overline \chi_{x+\hat\mu} \gamma_\mu \gamma_5 
\{A^R_{x,\mu},\chi_{x+\hat\mu}\}) \nonumber \\
&+&\frac{i F}{2} \sum_{x,\mu} \mbox{Tr}
(\overline \chi_x \gamma_\mu \gamma_5 [A^L_{x,\mu},\chi_{x}] +
\overline \chi_{x+\hat\mu} \gamma_\mu \gamma_5 
[A^R_{x,\mu},\chi_{x+\hat\mu}]) \nonumber \\
&+&\frac{r}{2} \sum_{x,\mu} \mbox{Tr} (2 \overline \chi_x \chi_x -
\overline \chi_x V_{x,\mu} \chi_{x+\hat\mu} V_{x,\mu}^\dagger -
\overline \chi_{x+\hat\mu} V_{x,\mu}^\dagger \chi_x V_{x,\mu})
\nonumber \\
&-&\frac{F_\pi^2}{4} \sum_{x,\mu} \mbox{Tr}[U_x^\dagger U_{x+\hat\mu} +
U_{x+\hat\mu}^\dagger U_x] - \frac{\langle \overline \psi \psi \rangle}{2}
\sum_x \mbox{Tr}[{\cal M} U_x^\dagger + {\cal M}^\dagger U_x].
\end{eqnarray}
The Wilson term proportional to $r$ removes the doubler fermions. In standard
lattice QCD this term breaks chiral symmetry explicitly. Remarkably, when
chiral symmetry is nonlinearly realized, not only the fermion mass term 
(proportional to $M$) but also the Wilson term is chirally invariant. The only
source of explicit chiral symmetry breaking is the current quark mass matrix
${\cal M}$. 

It is interesting to ask how the Nielsen-Ninomiya theorem \cite{Nie81} has been
avoided. Clearly, the actions (\ref{latact2}) and (\ref{latact3}) are local. 
For $N_f = 2$, the corresponding Dirac operator is given by
\begin{eqnarray}
\label{Dirac}
D_{xy}[U]&=&M \delta_{x,y} + 
\frac{1}{2} (\gamma_\mu V_{x,\mu} \delta_{x+\hat\mu,y} -
\gamma_\mu V_{x-\hat\mu,\mu}^\dagger \delta_{x-\hat\mu,y})
\nonumber \\
&+&\frac{i g_A}{2}
(\gamma_\mu \gamma_5 A^L_{x,\mu} \delta_{x,y} +
\gamma_\mu \gamma_5 A^R_{x-\hat\mu,\mu} \delta_{x,y})
\nonumber \\
&+&\frac{r}{2} (2 \delta_{x,y} - V_{x,\mu} \delta_{x+\hat\mu,y} -
V_{x-\hat\mu,\mu}^\dagger \delta_{x-\hat\mu,y}).
\end{eqnarray}
The Nielsen-Ninomiya theorem assumes that the Dirac operator anticommutes with
$\gamma_5$. This is not the case when chiral symmetry is nonlinearly realized, 
and hence one of the basic assumptions of the no-go theorem is not satisfied. 
Interestingly, Ginsparg-Wilson fermions evade the Nielsen-Ninomiya theorem by 
violating the same assumption. Of course, the nonlinear realization of chiral 
symmetry requires an explicit pion field which is not present in the 
fundamental QCD Lagrangian. 

Let us again discuss the issue of anomalies. On the lattice, anomalies may 
receive contributions from the doubler fermions. Hence, removing the doublers 
is intimately related to anomaly matching. It is straightforward to include 
electroweak gauge fields in the lattice theory as well. For example, the 
parallel transporters $W_{x,\mu} \in SU(2)_L$ of the electroweak gauge field 
transform as
\begin{equation}
W_{x,\mu}' = L_x W_{x,\mu} L_{x+\hat\mu}^\dagger.
\end{equation}
Together with the $U(1)_Y$ gauge field $B_{x,\mu}$, the field $W_{x,\mu}$ is
incorporated in the field
\begin{equation}
\tilde V_{x,\mu} = \frac{1}{2}[u_x^\dagger W_{x,\mu} u_{x+\hat\mu} +
u_x B_{x,\mu} u_{x+\hat\mu}^\dagger],
\end{equation}
which is again projected on a parallel transporter $V_{x,\mu} \in SU(2)$ by
a $GL(2)/SU(2)$ coset decomposition. Also the fields $A_{x,\mu}^L$ and
$A_{x,\mu}^R$ are constructed as in (\ref{Aleft}) and (\ref{Aright}), now using
\begin{equation}
\tilde A_{x,\mu} = \frac{i}{2}[u_x^\dagger W_{x,\mu} u_{x+\hat\mu} -
u_x B_{x,\mu} u_{x+\hat\mu}^\dagger].
\end{equation}
The lattice fermion action is constructed as in the ungauged case 
(\ref{latact2}) and is now $SU(2)_L \otimes U(1)_Y$ gauge invariant. 
Remarkably, not only the lattice fermion action but 
also the lattice fermion measure is gauge invariant. At first sight, this seems
to be a severe problem because gauge invariance of both the fermion action and
the fermion measure implies that lattice fermions with nonlinearly realized 
chiral symmetry do not contribute to anomalies. One might even suspect that the
doubler fermions have not been properly removed and thus have canceled the 
anomalies of the physical fermions. Fortunately, this is not the case. Indeed, 
as we have already seen in the continuum theory, fermions with a nonlinearly 
realized chiral symmetry do not contribute to anomalies. Instead, even in the 
presence of fermion fields, the anomalies are still contained in 
Goldstone-Wilczek and Wess-Zumino-Witten terms. Consequently, these terms 
should also be included in the lattice action. It is an interesting problem to 
construct lattice versions of Goldstone-Wilczek and Wess-Zumino-Witten terms 
maintaining the topological properties of the continuum theory. Since these 
terms do not play a central role in the applications discussed in the remainder
of the paper, we defer this problem to future investigations. It should be 
noted that the low-energy effective theory without the addition of 
Goldstone-Wilczek and Wess-Zumino-Witten terms, although consistent, does not 
correctly describe the low-energy physics of the underlying microscopic theory
with the anomalies. We also would like to point out that the topological terms 
give rise to complex action problems which may severely affect the efficiency 
of numerical simulations.

\section{Non-Perturbative Treatment of Low-Energy \\ Effective Theories on the
Lattice}

In this section we describe how non-perturbative calculations can be performed 
in low-energy effective theories on the lattice. In particular, we discuss how 
continuum physics can be extracted. As we have discussed in the introduction, 
non-perturbative problems indeed exist \cite{Leu87,Kap98,Bed98,Bed02} but have 
so far not required a lattice. This is likely to change when more complicated 
effective theories describing, for example, nuclear matter \cite{Mue00} are 
studied.

For simplicity, we first describe how the pure pion low-energy effective theory
can be treated on the lattice. It should be noted that, at present, there is no
need for such calculations. The only non-perturbative problem we are aware of 
in this context is the $SU(N_f)$ chiral rotor which has been solved 
analytically \cite{Leu87}. The leading term in the chiral Lagrangian for pions 
reads
\begin{equation}
\label{pupi_cont}
S[U] = \int d^4x \ 
\Big\{\frac{F_\pi^2}{4} \mbox{Tr}[\partial_\mu U^\dagger \partial_\mu U] -
\frac{\langle \overline \psi \psi \rangle}{2} \mbox{Tr}[{\cal M} U^\dagger +
{\cal M}^\dagger U]\Big\}.
\end{equation}
It is straightforward to discretize this theory. A simple lattice action reads
\begin{equation}
\label{pupi_latt}
S[U] = - \frac{F_\pi^2}{4} \sum_{x,\mu} \mbox{Tr}[U_x^\dagger U_{x+\hat\mu} +
U_{x+\hat\mu}^\dagger U_x] - \frac{\langle \overline \psi \psi \rangle}{2}
\sum_x \mbox{Tr}[{\cal M} U_x^\dagger + {\cal M}^\dagger U_x].
\end{equation}
Of course, the bare lattice parameters $F_\pi$ and 
$\langle \overline \psi \psi \rangle$ get renormalized. A basic statement of 
the low-energy effective theory is that, to leading order, the dynamics of 
pions depends only on the renormalized parameters $F_\pi^{\ren}$ and
$\langle \overline \psi \psi \rangle^{\ren}$. In the chiral limit, 
${\cal M} = 0$, the above lattice action has a phase transition at a critical 
bare coupling $F_{\pi c}$. For $N_f = 2$, the transition is second order, while
for $N_f \geq 3$, it is expected to be first order. For $F_\pi < F_{\pi c}$, 
one is in the symmetric phase without Goldstone bosons, while for 
$F_\pi > F_{\pi c}$, one is in the broken phase in which the Goldstone bosons
govern the low-energy physics. A non-perturbative calculation in the low-energy
effective theory can be performed anywhere in the broken phase. In particular, 
there is no need to approach a second order phase transition in order to 
extract continuum physics. This is because Goldstone bosons are naturally
massless without fine-tuning. However, the renormalized parameter 
$F_\pi^{\ren}$ will in general stay at the cut-off scale. Still, the low-energy
physics (which takes place far below the lattice cut-off) is universal and 
depends on the details of the lattice action only through $F_\pi^{\ren}$. In 
practical lattice calculations, one can choose any convenient bare coupling 
$F_\pi > F_{\pi c}$ in the broken phase and determine the corresponding 
renormalized coupling $F_\pi^{\ren}$. Then one can choose 
${\cal M} \langle \overline \psi \psi \rangle$ such that the 
physical pion mass $M_\pi^{\ren}$ is correctly reproduced (in units of 
$F_\pi^{\ren}$). After the low-energy parameters have been fixed to experiment,
any non-perturbative calculation of a low-energy phenomenon can be performed in
the lattice effective theory. Remarkably, at leading order, no lattice 
artifacts arise although $F_\pi^{\ren}$ is at the cut-off scale. Of course, in 
the lattice theory Lorentz invariance is reduced to discrete translations and 
rotations. Hence, there are terms in the low-energy theory of the lattice model
which are forbidden by Lorentz symmetry but not by the discrete rotation 
subgroup. Such terms are absent at leading order and arise only in the higher
orders of the low-energy expansion. Their prefactors are additional low-energy 
parameters (besides the usual Gasser-Leutwyler coefficients) which must be 
tuned such that, to the given order of the momentum expansion, the low-energy 
theory is free of lattice artifacts.

Non-perturbative calculations in the baryon effective theory can be performed 
in a similar way. The effective theory correctly describes the low-energy pion 
dynamics in the sector with one baryon. In this case, the bare parameters 
$M$ and $g_A$ (for $N_f = 2$) or $F$ and $D$ (for $N_f = 3$), as well as 
$F_\pi$ and ${\cal M} \langle \overline \psi \psi \rangle$ need to be tuned 
such that the corresponding renormalized
parameters match the experiment. Just like $F_\pi^{\ren}$, now also the baryon 
mass $M^{\ren}$ stays at the cut-off scale. This is no problem because at low 
energies the results of the effective theory are again universal and depend on 
the lattice details only through the values of the low-energy constants. There 
is no doubling problem as long as the mass of the doubler fermions is outside 
the energy range of the effective theory. In particular, there is no need to
approach a second order phase transition in order to make the physical baryon 
much lighter than the doubler fermions. Instead, by including higher order 
terms, one can push the doubler fermions to higher energies and thus extend the
validity range of the effective theory. More practical aspects of simulations 
of lattice theories with nonlinearly realized chiral symmetries will be 
discussed in the following sections.

\section{Static Baryons at Non-Zero Chemical Potential}

Nonlinear realizations of lattice chiral symmetry may also be useful beyond the
strict validity range of low-energy effective theories. In this 
section we discuss baryons in the static limit, $M \rightarrow \infty$, where 
interesting simplifications arise. In particular, one can then perform
simulations at non-zero baryon chemical potential without encountering a 
complex action problem. We first discuss the problem in the continuum and then
formulate the static baryon model on the lattice. Within this model we propose
exploratory studies of the QCD phase diagram at non-zero baryon chemical
potential.

In the continuum formulation, static baryons at an undetermined position 
$\vec x$ (with $g_A = 0$) are described by the spatial integral of the Polyakov
loop
\begin{equation}
\Phi[U] = \int d^3x \ \mbox{Tr} \ {\cal P} 
\exp[\int_0^\beta dt \ v_4(\vec x,t)]
\end{equation}
of the flavor ``gauge'' field $v_\mu(x)$ which is a function of 
$u(x)= U(x)^{1/2}$. Here ${\cal P}$ denotes path ordering and $\beta = 1/T$ is 
the inverse temperature. The partition function of pions in the background of a
single static baryon then takes the form
\begin{equation}
Z_B = \int {\cal D}U \ \Phi[U] \exp(- S[U]) \exp(- \beta M),
\end{equation}
where the baryon mass $M$ will ultimately be sent to infinity and
$S[U]$ is the pure pion action given in (\ref{pupi_cont}).
Similarly, the pion partition function in the 
presence of a single static antibaryon at an undetermined position is given by
\begin{equation}
Z_{\overline B} = \int {\cal D}U \ \Phi[U]^* \exp(- S[U]) \exp(- \beta M).
\end{equation}
In the next step, we consider a system of pions in the background of $n$ static
baryons and $\overline n$ static antibaryons which is described by the
partition function
\begin{equation}
Z_{n,\overline n} = \int {\cal D}U \ \frac{1}{n!} \Phi[U]^n 
\frac{1}{\overline n!} (\Phi[U]^*)^{\overline n} \exp(- S[U]) 
\exp[- \beta M (n + \overline n)].
\end{equation}
The permutation factors $n!$ and $\overline n!$ take into account the 
indistinguishability of baryons and antibaryons, respectively. Introducing the 
baryon chemical potential $\mu$ that couples to the baryon number 
$(n - \overline n)$, we obtain the grand canonical partition 
function
\begin{eqnarray}
Z(\mu)&=&\sum_{n,\overline n} Z_{n,\overline n} 
\exp[\beta \mu(n - \overline n)] \nonumber \\
&=&\sum_{n,\overline n} \int {\cal D}U \ \frac{1}{n!} \Phi[U]^n 
\frac{1}{\overline n!} (\Phi[U]^*)^{\overline n} 
\exp\{- S[U] - \beta n(M - \mu) - \beta \overline n(M + \mu)\} \nonumber \\
&=&\int {\cal D}U \ \exp\{- S[U] + \exp[- \beta (M - \mu)] \Phi[U] + 
\exp[- \beta (M + \mu)] \Phi[U]^*\}. \nonumber \\  \
\end{eqnarray}
The present calculation for baryons is consistent only for $M \rightarrow 
\infty$. Hence, in order to obtain a non-trivial result, we also send $\mu$ to
infinity keeping the difference $M - \mu$ finite. Then the partition function
reduces to
\begin{equation}
\label{partmu}
Z(\mu) = \int {\cal D}U \ \exp\{- S[U] + \exp[- \beta (M - \mu)] \Phi[U]\}.
\end{equation}
In the $N_f = 2$ case, the nucleons transform in the fundamental representation
of $SU(2)$ which has a real-valued trace. Consequently, the spatial integral of
the corresponding Polyakov loop
\begin{equation}
\Phi[U] = \int d^3x \ \mbox{Tr}_F {\cal P}
\exp[\int_0^\beta dt \ v_4(\vec x,t)],
\end{equation}
takes real values only. For $N_f = 3$, the trace in the fundamental 
representation is in general a complex number. However, since the baryons 
transform in the adjoint representation, the appropriate Polyakov loop averaged
over spatial positions reads
\begin{equation}
\Phi[U] = \int d^3x \ \mbox{Tr}_A {\cal P}
\exp[\int_0^\beta dt \ v_4(\vec x,t)] = \int d^3x \ \Big\{|\mbox{Tr}_F {\cal P}
\exp[\int_0^\beta dt \ v_4(\vec x,t)]|^2 - 1\Big\},
\end{equation}
which is again real. Hence, in contrast to full QCD, both for $N_f = 2$ and for
$N_f = 3$ no complex action problem arises. Using similar methods, the
thermodynamics of gluons in the background of static quarks have been
investigated \cite{Ben92,Blu96,Eng99}. In that case, a complex action problem 
does arise but, at least in the Potts model approximation to QCD, could be 
solved with a meron-cluster algorithm \cite{Alf01}.

It is straightforward to formulate the static baryon model on the lattice. For 
$N_f = 2$, the spatial integral of the Polyakov loop for a static nucleon is 
given by
\begin{equation}
\Phi[U] = \sum_{\vec x} \mbox{Tr}_F \prod_t V_{\vec x,t,4},
\end{equation}
and, for $N_f = 3$, the corresponding quantity for a static baryon takes the 
form
\begin{equation}
\Phi[U] = \sum_{\vec x} \mbox{Tr}_A[\prod_t V_{\vec x,t,4}] =
\sum_{\vec x} \Big\{|\mbox{Tr}_F \prod_t V_{\vec x,t,4}|^2 - 1\Big\}.
\end{equation}
Then, the lattice partition function corresponding to (\ref{partmu}) reads
\begin{equation}
Z(\mu) = \prod_x \int_{SU(N_f)} dU_x \ 
\exp\{- S[U] + \exp[- \beta (M - \mu)] \Phi[U]\}
\end{equation}
with the pure pion lattice action $S[U]$ given in (\ref{pupi_latt}).
It should be noted that we have ignored the Pauli principle for static baryons
occupying the same lattice site. The effect of this approximation is a lattice 
artifact that does not affect the continuum limit.

The static baryon model can be used in exploratory studies of the QCD phase 
diagram at non-zero chemical potential. At $\mu = 0$, the pure 
pion lattice model has a finite temperature chiral phase transition, above 
which chiral symmetry is restored. For two massless flavors, this transition is
second order and it is in the universality class of the 3-dimensional 
$O(4)$-model. This transition extends into the $(\mu,T)$-plane and is expected 
to turn into a first order phase transition at a tricritical point. When 
non-zero current quark masses are included, the second order phase transition 
line is washed out to a crossover and the tricritical point turns into a 
critical endpoint of the remaining line of first order phase transitions. 
Locating the critical endpoint in the $(\mu,T)$-plane is an important issue in 
heavy ion collision experiments \cite{Ste98} which has been addressed using 
lattice methods \cite{Fod02}. Unfortunately, in these studies one is limited to
small lattices due to a severe complex action problem. The small-$\mu$ region 
has also been investigated in \cite{All02,deF02}. One should not expect to be 
able to extract the precise location of the critical endpoint using the static 
baryon model presented here. However, as long as the model has the correct 
symmetry properties, one can at least investigate universal features of this 
point, which is expected to be in the universality class of the 3-dimensional 
Ising model. Moreover, one can gain experience with numerical methods for 
locating the critical endpoint in a model that is similar to full QCD but much 
simpler to simulate.

We would like to mention that the static baryon model shares all global
symmetries with QCD, but is, in addition, also invariant under the 
intrinsic parity transformation $P_0$. In the $N_f = 2$ case, $P_0$ is just
$G$-parity and hence a symmetry of QCD. In the $N_f = 3$ case, $P_0$ is not a 
symmetry of QCD. At the level of the effective theory, this symmetry is 
explicitly broken by the Wess-Zumino-Witten term in the Goldstone boson sector
and by the $g_A$ term in the baryon sector. Since we have not included the 
Wess-Zumino-Witten term or the $g_A$ term in the lattice action, as it stands, 
the $N_f = 3$ static baryon model has the undesirable $P_0$ symmetry. In a 
study of universal properties of the $N_f = 3$ model one may hence want to 
break $P_0$ explicitly by hand in order to exactly reproduce all symmetries of 
QCD.

With light up and down quarks and a very heavy strange quark, the
situation is essentially like in the $N_f = 2$ case. As one lowers the strange
quark mass, the critical endpoint is expected to move toward smaller values of
$\mu$ until it reaches the temperature axis. For even smaller strange quark
masses, the chiral phase transition at $\mu = 0$ is first order. Lattice
calculations suggest (but do not show unambiguously) that, for Nature's quark 
mass values, at $\mu = 0$ there is merely a crossover. In that case, the 
critical endpoint would indeed be located in the $(\mu,T)$-plane at $\mu > 0$.
While one cannot determine the precise location of the critical endpoint in QCD
from the static baryon model, it should be possible to obtain qualitative (and 
perhaps even semi-quantitative) information concerning the location of this
endpoint as a function of the strange quark mass.

\section{Constituent Quarks on the Lattice}

In this section we use a nonlinearly realized lattice chiral symmetry in order
to address non-perturbative questions concerning the confinement of constituent
quarks. Again, we do not rely on the validity of a systematic low-energy 
expansion. The chiral quark model of Georgi and Manohar \cite{Geo84b} offers an
explanation for why the non-re\-la\-ti\-vis\-tic quark model works. However, 
there are some potential non-perturbative problems which we suggest to 
investigate on the lattice.

The chiral quark model is formulated in terms of gluons, pions, and constituent
quarks $\Psi(x)$ and $\overline \Psi(x)$ with the Euclidean action
\begin{eqnarray}
S[\overline \Psi,\Psi,U,G_\mu]&=&\int d^4x \
\Big\{M \overline \Psi \Psi + 
\overline \Psi \gamma_\mu (\partial_\mu + v_\mu + G_\mu) \Psi +
i g_A \overline \Psi \gamma_\mu \gamma_5 a_\mu \Psi \nonumber \\
&+&\frac{F_\pi^2}{4} \mbox{Tr}[\partial_\mu U^\dagger \partial_\mu U] -
\frac{\langle \overline \psi \psi \rangle}{2} \mbox{Tr}[{\cal M} U^\dagger +
{\cal M}^\dagger U] \nonumber \\
&+&\frac{1}{2 g_s^2} \mbox{Tr}[G_{\mu\nu} G_{\mu\nu}]\Big\}.
\end{eqnarray}
Here $G_\mu(x) = i g_s G_\mu^a(x) \lambda^a$ is the $SU(N_c)$ gluon field, 
and $g_s$ is the strong coupling constant between gluons and constituent quarks
(transforming in the fundamental representations of both $SU(N_c)$ color and 
$SU(N_f)$ flavor). Under gauge transformations $g(x) \in SU(N_c)$ the gluon 
field transforms as 
\begin{equation}
G_\mu(x)' = g(x)[G_\mu(x) + \partial_\mu] g(x)^\dagger.
\end{equation}
As usual, the field strength takes the form
\begin{equation}
G_{\mu\nu}(x) = \partial_\mu G_\nu(x) - \partial_\nu G_\mu(x) + 
[G_\mu(x),G_\nu(x)].
\end{equation}
We limit ourselves to $N_c \geq 3$. The $N_c = 2$ case is exceptional because 
$SU(2)$ is pseudo-real and hence the pattern of chiral symmetry breaking is 
different \cite{Pes80}. 

The chiral quark model of Georgi and Manohar is based 
on the assumption that the energy scale for chiral symmetry breaking is larger 
than the one for confinement. An effective description in terms of constituent 
quarks which receive their mass from chiral symmetry breaking before they get 
confined by residual low-energy gluons should then make sense. In fact, the 
phenomenological success of the non-relativistic quark model may suggest that 
this picture is indeed correct. In their work, Georgi and Manohar provided a
framework that puts the non-relativistic quark model on a solid field 
theoretical basis. In particular, their chiral quark model holds the promise to
share the successes of the non-relativistic quark model. For example, making
reasonable assumptions about the confinement of constituent quarks, it 
describes baryon magnetic moments and hyperon non-leptonic decays with about 10
percent accuracy. In addition, in contrast to the non-relativistic quark model,
the Georgi-Manohar model is manifestly chirally invariant and is thus 
automatically consistent with the predictions of chiral perturbation theory.

Georgi and Manohar also pointed out some potential problems of their model. For
example, the constituent quarks may form a pseudo-scalar bound state in 
addition to the explicit pion introduced as a fundamental field. If this bound 
state is light, one would double-count the pion. Georgi and Manohar argued that
this should not happen. The pseudo-scalar constituent quark bound state should 
either be so heavy that it lies outside the range of the low-energy theory or 
it might represent the next physical pseudo-scalar state above the pion. A 
similar issue arises for the baryons. In addition to the baryons built from 
constituent quarks, one can form baryonic Skyrmions from the fundamental pion 
field. Again, Georgi and Manohar argued that double-counting should not arise 
because the Skyrmions may either be very heavy or unstable. A potentially more 
severe problem is related to the confinement scale. In the chiral quark model 
the strong coupling constant is fixed by adjusting the color hyperfine 
splitting in the baryon spectrum. The estimated value of $g_s$ is so small that
one might expect unacceptably low-lying glueball states. For the same reason, 
there might be a low-temperature deconfinement phase transition in the gluon 
sector significantly below the finite temperature chiral phase transition.

These potential problems are impossible to address quantitatively in the
continuum formulation of the chiral quark model because they involve the
non-perturbative dynamics of confinement. In order to be able to address these 
issues, we formulate the constituent quark model on the lattice. Due to the 
nonlinear realization of chiral symmetry, global axial transformations manifest
themselves as pion field dependent (and thus local) transformations on the 
constituent quark fields. In particular, the constituent quarks couple to the 
pions through the composite flavor ``gauge'' field. 

Using the lattice construction of a nonlinearly realized chiral symmetry 
presented in section 3, it is now straightforward to put the Georgi-Manohar
model on the lattice. The resulting action takes the form
\begin{eqnarray}
\label{GMaction}
S[\overline \Psi,\Psi,U,G]&=&M \sum_x \overline \Psi_x \Psi_x + 
\frac{1}{2} \sum_{x,\mu} 
(\overline \Psi_x \gamma_\mu V_{x,\mu} G_{x,\mu} \Psi_{x+\hat\mu} -
\overline \Psi_{x+\hat\mu} \gamma_\mu V_{x,\mu}^\dagger G_{x,\mu}^\dagger
\Psi_x) \nonumber \\
&+&\frac{i g_A}{2} \sum_{x,\mu}  
(\overline \Psi_x \gamma_\mu \gamma_5 A^L_{x,\mu} \Psi_{x} +
\overline \Psi_{x+\hat\mu} \gamma_\mu \gamma_5 A^R_{x,\mu} \Psi_{x+\hat\mu})
\nonumber \\
&+&\frac{r}{2} \sum_{x,\mu} (2 \overline \Psi_x \Psi_x -
\overline \Psi_x V_{x,\mu} G_{x,\mu} \Psi_{x+\hat\mu} -
\overline \Psi_{x+\hat\mu} V_{x,\mu}^\dagger G_{x,\mu}^\dagger \Psi_x)
\nonumber \\
&-&\frac{F_\pi^2}{4} \sum_{x,\mu} \mbox{Tr}[U_x^\dagger U_{x+\hat\mu} +
U_{x+\hat\mu}^\dagger U_x] - \frac{\langle \overline \psi \psi \rangle}{2}
\sum_x \mbox{Tr}[{\cal M} U_x^\dagger + {\cal M}^\dagger U_x] \nonumber \\
&-&\frac{1}{g_s^2} \sum_{x,\mu,\nu} \mbox{Tr}
[G_{x,\mu} G_{x+\hat\mu,\nu} G_{x+\hat\nu,\mu}^\dagger G_{x,\nu}^\dagger +
G_{x,\nu} G_{x+\hat\nu,\mu} G_{x+\hat\mu,\nu}^\dagger G_{x,\mu}^\dagger].
\end{eqnarray}
Here $G_{x,\mu} \in SU(N_c)$ (again taking $N_c \geq 3$) denotes the standard
color parallel transporters living on the lattice links, which transform under
gauge transformations $g_x \in SU(N_c)$ as
\begin{equation}
G_{x,\mu}' = g_x G_{x,\mu} g_{x+\hat\mu}^\dagger.
\end{equation}
In the classical continuum limit the lattice gauge field
$G_{x,\mu} = \exp[G_\mu(x+\hat\mu)]$ is related to the continuum gauge field 
$G_\mu(x)$ that lives in the $su(N_c)$ algebra.


For $g_A = 0$, the Dirac operator of the chiral quark model
\begin{eqnarray}
D_{xy}[U,G]&=&M \delta_{x,y} + 
\frac{1}{2} (\gamma_\mu V_{x,\mu} G_{x,\mu} \delta_{x+\hat\mu,y} -
\gamma_\mu V_{x-\hat\mu,\mu}^\dagger G_{x-\hat\mu,\mu}^\dagger 
\delta_{x-\hat\mu,y})
\nonumber \\
&+&\frac{i g_A}{2}
(\gamma_\mu \gamma_5 A^L_{x,\mu} \delta_{x,y} +
\gamma_\mu \gamma_5 A^R_{x-\hat\mu,\mu} \delta_{x,y})
\nonumber \\
&+&\frac{r}{2} (2 \delta_{x,y} - V_{x,\mu} G_{x,\mu} \delta_{x+\hat\mu,y} -
V_{x-\hat\mu,\mu}^\dagger G_{x-\hat\mu,\mu}^\dagger \delta_{x-\hat\mu,y}),
\end{eqnarray}
is $\gamma_5$-Hermitean, i.e. $D[U,G]^\dagger = \gamma_5 D[U,G] \gamma_5$. 
Thus, its spectrum contains complex conjugate pairs of eigenvalues $\lambda$ 
and $\lambda^*$. Only the real eigenvalues are in general not paired. 
Consequently, the fermion determinant can be negative only if there is an odd 
number of negative (real and thus unpaired) eigenvalues. In the presence of a 
sufficiently large constituent quark mass $M$, negative eigenvalues are very 
much suppressed. Hence, the potential sign problem due to a negative fermion 
determinant is expected to be mild. A similar situation arises when one wants 
to simulate an odd number of flavors of Wilson fermions, for example, 
when one wants to include the strange quark in a QCD simulation. In this case 
--- not directly addressing the potential sign problem --- algorithms have been
developed to perform numerical simulations \cite{Bor95,Ale99,Tak01,Aok02}. It 
is to be expected that similar methods can be applied to the action presented 
here. Unfortunately, for $g_A \neq 0$ the Dirac operator is in general no 
longer $\gamma_5$-Hermitean. Hence, we can no longer argue that there is only a
mild sign problem. Only numerical simulations can tell if a severe sign problem
arises for realistic values of $g_A$. If this is the case, one might want to 
simulate at purely imaginary values of $g_A$ and attempt an analytic 
continuation to the real physical value. However, in order to address the
non-perturbative questions concerning the confinement of constituent quarks, it
may be sufficient to work at $g_A = 0$.

In the chiral limit (${\cal M} = 0$) and for $r = 1$ the lattice Georgi-Manohar
model has four bare parameters: the constituent quark mass $M$, the axial 
coupling $g_A$, the pion decay constant $F_\pi$, and the strong gauge coupling 
$g_s$. The chiral quark model on the lattice can be treated similarly to the
low-energy baryon effective theory. In particular, one need not necessarily
approach a second order phase transition but can simulate anywhere in the 
chirally broken phase. For example, one can tune the bare constituent quark 
mass $M$ and the bare pion decay constant $F_\pi$ such that both the 
renormalized nucleon mass $M_{\mathrm{N}}^{\ren}$ and the renormalized pion 
decay constant $F_\pi^{\ren}$ stay at the cut-off with their ratio equal to the
experimental value. Similarly, the bare $g_A$ can be tuned to reproduce the 
observed weak decay of the neutron and $g_s$ can be adjusted to the 
nucleon-delta mass splitting. With the bare parameters fixed in this way, one 
can ask if the low-energy physics in the gluon sector is consistent with 
experiment. For example, it is interesting to investigate if there is a first 
order deconfinement phase transition at an unusually low temperature or if 
there are unacceptably light glueball states. If potential problems of this 
kind do not arise, the chiral quark model may indeed be able to explain the 
phenomenological success of the non-relativistic quark model. 

Unlike the baryon effective theory, we do not expect the chiral quark model to 
represent a systematic low-energy expansion. In particular, one should not 
expect that its low-energy gluon physics is universal and depends on $M$, 
$g_A$, $F_\pi$, and $g_s$ only. Still, despite the fact that other details of 
the model are expected to influence the gluon sector, it is interesting to ask 
if the gluon dynamics can be modeled successfully in this way. Since one cannot
expect to obtain precise results directly relevant to QCD, one may ask if using
the machinery of lattice field theory is at all justified. Clearly, this is a 
matter of taste. However, even if only qualitative insight into the success of 
the non-relativistic quark model can be gained, this may be interesting enough.
In addition, a lattice study of the Georgi-Manohar model would also provide 
valuable experience with dynamical fermion algorithms with exact chiral 
symmetry in a set-up that is much simpler than lattice QCD with Ginsparg-Wilson
fermions.

\section{From the Chiral Quark Model to QCD?}

It is interesting to ask if a nonlinearly realized chiral symmetry can also 
lead to a new approach to lattice QCD itself. Using a linear realization of
chiral symmetry and an explicit auxiliary pion field, similar questions have 
been addressed in \cite{Bro94,Kog98}. In this section we investigate such 
questions by exploring the phase structure of the lattice chiral quark model as
a function of its bare parameters $M$, $F_\pi$, and $g_s$. For simplicity, we 
put $g_A = 0$ and $r = 1$. 

In the limit of infinite constituent quark mass, $M = \pm \infty$, the pions 
and gluons decouple from each other: the model splits into an $SU(N_c)$ 
Yang-Mills theory and an $SU(N_f)_L \times SU(N_f)_R$ nonlinear $\sigma$-model 
for the pions. The pure gluon theory is expected to have no bulk phase 
transition in the bare gauge coupling $g_s$ and, as a consequence of asymptotic
freedom, reaches its continuum limit as $g_s \rightarrow 0$. For large values 
of $F_\pi^2$, the pion field $U_x$ is ordered 
and chiral symmetry is spontaneously broken to $SU(N_f)_{L=R}$, while for small
values of $F_\pi^2$, it is restored. At $M = \pm \infty$ the critical coupling 
$F_{\pi c}^2$ is independent of $g_s$. Consequently, there is a horizontal line
connecting the points $A$ and $B$ in the phase diagram of fig.~1. In the 
$N_f = 2$ case, the phase transition is second order, while for $N_f = 3$, it 
is expected to be first order.
\begin{figure}[htb]
\begin{center}
\epsfig{figure=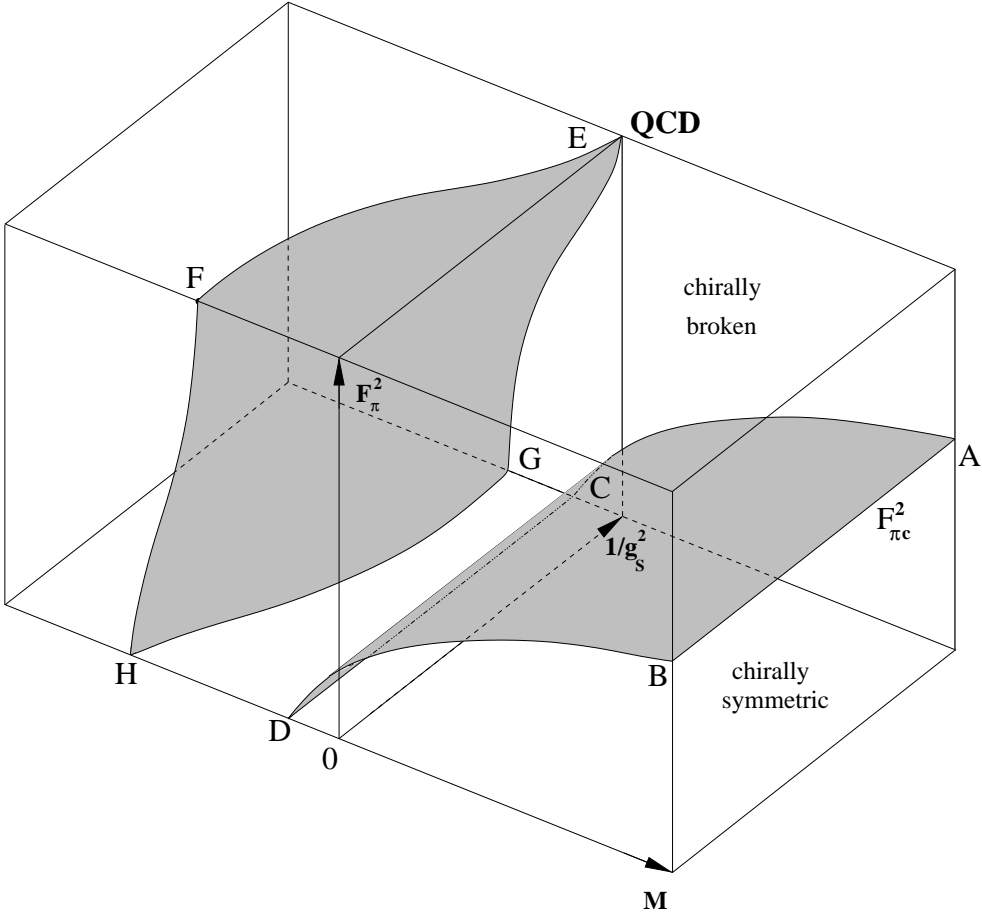,width=13cm}
\end{center}
\caption{\it Possible form of the phase diagram of the chiral quark model in 
the $(M,F^2_\pi,1/g_s^2)$-space at $g_A = 0$ and $r = 1$. Structures behind the
critical surface connecting the points $E$, $F$, $G$, and $H$ are not shown.
The ranges of the variables are $M \in [-\infty,\infty]$ and $F_\pi^2, 1/g_s^2
\in [0,\infty]$.}
\end{figure}

Next we consider the limit of infinite bare coupling $F_\pi^2 = \infty$. The 
pion field $U_x = u_x^2$ is then frozen to a space-time independent constant 
$U_0$. Of course, this does not mean that $u_x$ is a constant. This can be 
illustrated as follows. If $U_x = U_0$ for all $x$, all $u_x$ can be 
diagonalized by the same unitary transformation $W$, i.e. 
$u_x = W d_x W^\dagger$. The matrix $d_x = d s_x$ is a constant diagonal matrix
$d \in U(1)^{N_f-1}$ up to a diagonal matrix $s_x \in \Z(2)^{N_f-1}$ obeying 
$s_x^2 = \1$. The Hermitean matrices $s_x$ have diagonal elements $\pm 1$ and 
determinant 1. Hence, we can now write
\begin{equation}
u_x = W d s_x W^\dagger,
\end{equation}
such that indeed $u_x^2 = W d^2 W^\dagger = U_0$ is independent of $x$. Let us
now consider the other fields in the $F_\pi^2 = \infty$ limit. The lattice 
field $\tilde V_{x,\mu}$ takes the form
\begin{equation}
\tilde V_{x,\mu} = \frac{1}{2}[u_x^\dagger u_{x+\hat\mu} + 
u_x u_{x+\hat\mu}^\dagger] = W s_x s_{x+\hat\mu} W^\dagger = V_{x,\mu}.
\end{equation}
Since $\tilde V_{x,\mu}$ is already in $SU(N_f)$, the $GL(N_f)/SU(N_f)$ coset 
decomposition yields $V_{x,\mu} = \tilde V_{x,\mu}$. Similarly, it is 
straightforward to show that $\tilde A_{x,\mu} = A_{x,\mu}^L = A_{x,\mu}^R = 
0$. We now perform the transformation $V_x = s_x W^\dagger$ on the constituent 
quark field
\begin{equation}
\Psi_x = V_x^\dagger \Psi_x' =  W s_x \Psi_x', \ \overline \Psi_x = 
\overline \Psi_x' V_x = \overline \Psi_x s_x W^\dagger.
\end{equation}
As a consequence,
\begin{eqnarray}
&&\overline \Psi_x \Psi_x = \overline \Psi_x' s_x W^\dagger W s_x \Psi_x' =
\overline \Psi_x' \Psi_x', \nonumber \\
&&\overline \Psi_x V_{x,\mu} \Psi_{x+\hat\mu} = 
\overline \Psi_x' s_x W^\dagger W s_x s_{x+\hat\mu} W^\dagger W s_{x+\hat\mu}
\Psi_{x+\hat\mu}' = \overline \Psi_x' \Psi_{x+\hat\mu}', \nonumber \\
&&\overline \Psi_{x+\hat\mu} V_{x,\mu}^\dagger \Psi_x = 
\overline \Psi_{x+\hat\mu}' s_{x+\hat\mu} W^\dagger W s_{x+\hat\mu} s_x 
W^\dagger W s_x \Psi_x' = \overline \Psi_{x+\hat\mu}' \Psi_x',
\end{eqnarray}
so that (for $F_\pi^2 = \infty$) the action (\ref{GMaction}) takes the form
\begin{eqnarray}
S[\overline \Psi',\Psi',G]&=&M \sum_x \overline \Psi_x' \Psi_x' + 
\frac{1}{2} \sum_{x,\mu} 
(\overline \Psi_x' \gamma_\mu G_{x,\mu} \Psi_{x+\hat\mu}' -
\overline \Psi_{x+\hat\mu}' \gamma_\mu G_{x,\mu}^\dagger \Psi_x')
\nonumber \\
&+&\frac{r}{2} \sum_{x,\mu} (2 \overline \Psi_x' \Psi_x' -
\overline \Psi_x' G_{x,\mu} \Psi_{x+\hat\mu}' -
\overline \Psi_{x+\hat\mu}' G_{x,\mu}^\dagger \Psi_x')
\nonumber \\
&-&\frac{1}{g_s^2} \sum_{x,\mu,\nu} \mbox{Tr}
[G_{x,\mu} G_{x+\hat\mu,\nu} G_{x+\hat\nu,\mu}^\dagger G_{x,\nu}^\dagger +
G_{x,\nu} G_{x+\hat\nu,\mu} G_{x+\hat\mu,\nu}^\dagger G_{x,\mu}^\dagger].
\nonumber \\ \
\end{eqnarray}
Remarkably, this is nothing but the standard Wilson fermion action. Hence, in
the limit $F_\pi^2 = \infty$, $g_s \rightarrow 0$, and $M \rightarrow 0$ (point
$E$ in fig.~1), we reach the usual continuum limit of lattice QCD. This point 
is connected to the point $F$ at strong coupling ($g_s = \infty$) by a line of 
second order phase transitions. Fine-tuning to this line, a massless ``pion'' 
emerges as a quark-antiquark bound state. This ``pion'' is not the Goldstone 
boson that results from chiral symmetry breaking, except in the continuum 
limit. Still, in Wilson's approach to lattice QCD this particle represents the 
physical pion. Then, recovering chiral symmetry requires fine-tuning as well as
taking the continuum limit. Of course,
in the chiral quark model with a nonlinearly realized chiral symmetry, there is
a genuine Goldstone pion. However, this particle decouples at $F_\pi^2 = 
\infty$. It is tempting to try to avoid fine-tuning and to represent the 
physical pion by the Goldstone boson of the nonlinearly realized chiral 
symmetry. This would suggest to approach the QCD point $E$ from finite values 
of $F_\pi^2$ out of the chirally broken phase. As one approaches the point $E$,
one should then avoid the critical surface connecting the points $E$, $F$, $G$,
and $H$ because otherwise one would double-count the pion. However, it is not 
clear if one will obtain a new approach to QCD in this way. For example, it is 
likely that the explicit Goldstone pion decouples and that the QCD pion must 
still be assembled as a quark-antiquark bound state. 

Let us now consider the $g_s = 0$ plane in the phase diagram. Then the gluon
field is completely frozen to a pure gauge $G_{x,\mu} = 
g_x g_{x+\hat\mu}^\dagger$. It should be noted that the continuum limit of 
lattice QCD is taken by sending $g_s \rightarrow 0$ in a controlled way, not
by simply putting $g_s = 0$. In particular, at $g_s = 0$ there is no gluon
field and thus no confinement. In this case, we transform the fermion field as
$\Psi_x = g_x  \Psi_x'$, $\overline \Psi_x = \overline \Psi_x' g_x^\dagger$, 
so that
\begin{eqnarray}
&&\overline \Psi_x \Psi_x = \overline \Psi_x' g_x^\dagger g_x \Psi_x' = 
\overline \Psi_x' \Psi_x', \nonumber \\
&&\overline \Psi_x G_{x,\mu} \Psi_{x+\hat\mu} = 
\overline \Psi_x' g_x^\dagger g_x g_{x+\hat\mu}^\dagger g_{x+\hat\mu} 
\Psi_{x+\hat\mu}' = \overline \Psi_x' \Psi_{x+\hat\mu}', \nonumber \\
&&\overline \Psi_{x+\hat\mu} G_{x,\mu}^\dagger \Psi_x = 
\overline \Psi_{x+\hat\mu}' g_{x+\hat\mu}^\dagger g_{x+\hat\mu} g_x^\dagger g_x
\Psi_x' = \overline \Psi_{x+\hat\mu}' \Psi_x'.
\end{eqnarray}
Hence, for $g_s = 0$ the action (\ref{GMaction}) reduces to
\begin{eqnarray}
\label{lataction}
S[\overline \Psi',\Psi',U]&=&M \sum_x \overline \Psi_x' \Psi_x' + 
\frac{1}{2} \sum_{x,\mu}
(\overline \Psi_x' \gamma_\mu V_{x,\mu} \Psi_{x+\hat\mu}' -
\overline \Psi_{x+\hat\mu}' \gamma_\mu V_{x,\mu}^\dagger \Psi_x') \nonumber \\
&+&\frac{i g_A}{2} \sum_{x,\mu}
(\overline \Psi_x' \gamma_\mu \gamma_5 A^L_{x,\mu} \Psi_{x}' +
\overline \Psi_{x+\hat\mu}' \gamma_\mu \gamma_5 A^R_{x,\mu} \Psi_{x+\hat\mu}')
\nonumber \\
&+&\frac{r}{2} \sum_{x,\mu} (2 \overline \Psi_x' \Psi_x' -
\overline \Psi_x' V_{x,\mu} \Psi_{x+\hat\mu}' -
\overline \Psi_{x+\hat\mu}' V_{x,\mu}^\dagger \Psi_x')
\nonumber \\
&-&\frac{F_\pi^2}{4} \sum_{x,\mu} \mbox{Tr}[U_x^\dagger U_{x+\hat\mu} +
U_{x+\hat\mu}^\dagger U_x] - \frac{\langle \overline \psi \psi \rangle}{2}
\sum_x \mbox{Tr}[{\cal M} U_x^\dagger + {\cal M}^\dagger U_x], \nonumber \\ \
\end{eqnarray}
which has exactly the same form as the baryon action (\ref{latact2}). Of 
course, here the field $\Psi_x$ describes a colored constituent quark. Since we
have now switched off the strong interactions, these quarks are no longer 
confined. 

\begin{figure}[htb]
\begin{center}
\epsfig{figure=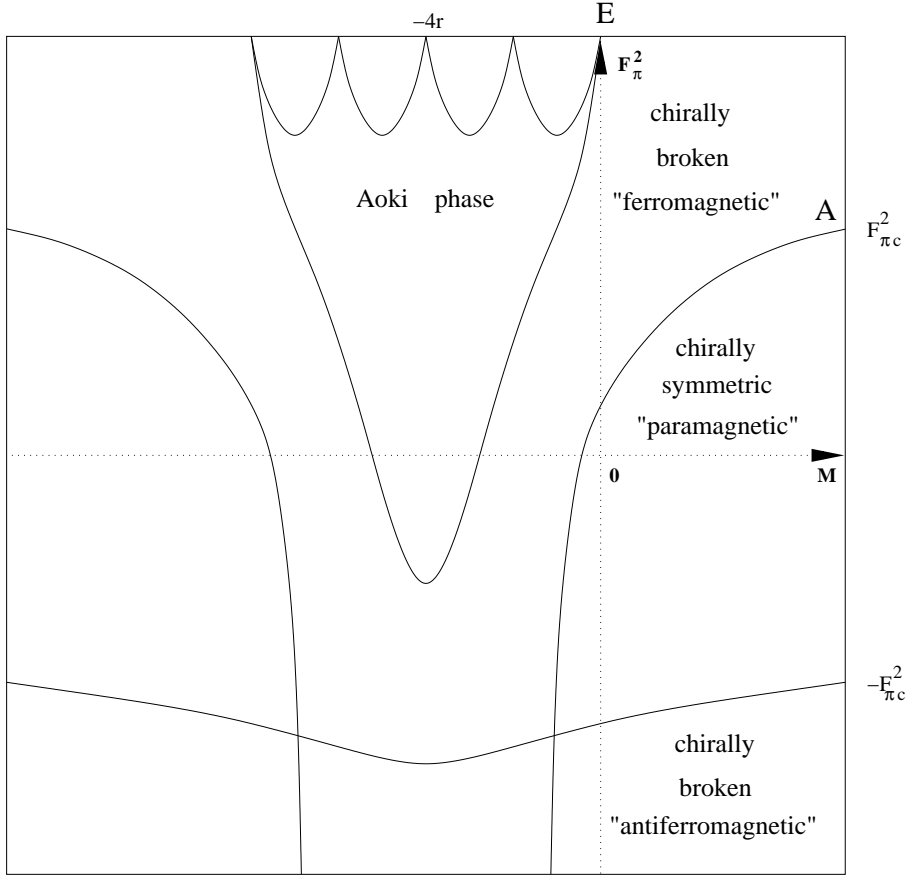,width=12cm}
\end{center}
\caption{\it Possible form of the phase diagram of the chiral quark model in 
the $(M,F_\pi^2)$-plane at $g_s = 0$, $g_A = 0$, and $r = 1$. The ranges of the
variables are $M, F_\pi^2 \in [-\infty,\infty]$.}
\end{figure}
For $N_f = 2$, a possible form of the phase diagram in the $(M,F_\pi^2)$-plane 
(at $g_s = 0$) is depicted in fig.~2. It also includes the Aoki phase
not shown in fig.~1. In the limit of infinite constituent quark mass, 
$M = \pm \infty$, the model reduces to an $SU(2)_L \otimes SU(2)_R = O(4)$ 
nonlinear $\sigma$-model for the pions. For $F_\pi^2 > F_{\pi c}^2$, the pion 
field $U_x$ is ordered and hence chiral symmetry is spontaneously broken to 
$SU(2)_{L=R} = O(3)$, while for $F_\pi^2 < F_{\pi c}^2$, it is restored. It is 
interesting to consider negative values of $F_\pi^2$ as well (still keeping 
$M = \pm \infty$). On a bi-partite lattice one can map the theory back to the 
one with positive values of $F_\pi^2$ by changing the sign of $U_x$ on all odd 
lattice sites. Hence, for $F_\pi^2 < - F_{\pi c}^2$, there is an 
``antiferromagnetically'' ordered phase in which chiral symmetry is again 
broken from $O(4)$ to $O(3)$. Hence, for $M = \pm \infty$, there are two 
phase transitions, one at $F_{\pi c}^2$ separating the usual ``ferromagnetic''
chirally broken from the ``paramagnetic'' chirally symmetric phase and another 
one at $- F_{\pi c}^2$ separating the ``paramagnetic'' from the 
``antiferromagnetic'' broken phase.

At $M = 0$ and $F_\pi^2 = \infty$ (point $E$ in figs.~1 and 2) there is a 
second order phase transition at which the constituent quark mass vanishes. In 
addition, 
there are unphysical second order phase transitions at $M + 2 n r = 0$ (with 
$n \in \{1,2,3,4\}$) at which some of the doubler fermions become massless. In
the context of Wilson's lattice QCD the region of $M < 0$ was explored by
Aoki \cite{Aok84}. He identified a rich structure including a phase in which 
parity and flavor are spontaneously broken. It should be noted that for 
$g_A = 0$ the model (\ref{lataction}) has a symmetry under the replacement of 
$M$ by $- (M + 4 r)$. This replacement can be compensated for by changing the 
sign of $\Psi_x$ on all even sites and the sign of $\overline \Psi_x$ on all 
odd sites. This symmetry manifests itself in the phase diagram of fig.~2. Our 
sketch of the Aoki phase in fig.~2 is inspired by its form in Wilson's lattice 
QCD. The precise structure of the phase diagram can only be inferred from 
numerical simulations. The phase diagram of fig.~2 is a possible interpolation 
between the limits $M = \pm \infty$ and $F_\pi^2 = \infty$, partly inspired by 
the phase diagram of the Smit-Swift model and other related Higgs-Yukawa models
\cite{Smi80,Swi84,Boc90}. Attempts to construct chiral gauge theories as the 
continuum limit of such lattice models have failed 
because it was impossible to remove the doubler fermions, both at weak and at 
strong Yukawa couplings \cite{Pet93}. Such problems do not affect our 
construction of lattice effective theories with nonlinearly realized chiral 
symmetry because, in that case, there is no need to approach a second order 
phase transition to extract low-energy continuum physics. However, attempts to 
construct full QCD with a manifest nonlinearly realized chiral symmetry on the 
lattice are likely to fail for the same reasons as the Smit-Swift model did. 
Still, it is interesting to ask if the lattice chiral symmetry of 
Ginsparg-Wilson fermions can be realized nonlinearly. We leave such questions 
for future investigations.

\section{Conclusions}

In this paper we have constructed lattice theories with baryon or constituent 
quark fields on which chiral symmetry is nonlinearly realized. In this 
framework the fermion doublers are removed by the Wilson term without breaking 
chiral symmetry explicitly. This is possible because fermion fields do not 
contribute to anomalies in theories with nonlinearly realized chiral 
symmetries. Instead the anomalies must be included explicitly in the form of 
Wess-Zumino-Witten and Goldstone-Wilczek terms. The lattice construction of
such terms remains an interesting challenge for future investigations.

The use of nonlinearly realized chiral symmetry is well established in the
context of chiral perturbation theory which provides a power counting scheme
for a systematic low-energy expansion. However, also non-perturbative problems 
may arise in the low-energy regime. Some of these problems can be dealt with
analytically \cite{Leu87}. In general, however, such problems require some
numerical work. While one can still proceed without Monte Carlo methods in
investigations of few-nucleon systems \cite{Bed98,Bed02}, this is likely to 
change as one moves on to larger nuclei and ultimately to nuclear matter 
\cite{Mue00}. Our lattice formulation provides a framework in which future 
numerical studies of this kind can be performed on a solid theoretical basis.

Even if one leaves the framework of a systematic low-energy expansion, the
nonlinear realization of chiral symmetry on the lattice may provide insight
into interesting physical questions. For example, as we have seen, in a model
of static baryons at non-zero chemical potential one can investigate universal
aspects of the QCD phase diagram, like the nature of the critical endpoint of
the chiral phase transition line. In the static baryon approximation, one 
cannot expect to obtain results directly relevant to QCD for non-universal 
quantities such as the location of the critical endpoint. Still, one may be 
able to collect, for example, qualitative information on how the critical 
endpoint moves as the strange quark mass is varied.

Another interesting example is the chiral quark model of Georgi and Manohar. It
holds the promise to explain the success of the non-relativistic quark model
and has successfully described a variety of physical properties of hadrons.
However, there is a number of potential dynamical problems that can only be
addressed non-perturbatively. Our lattice formulation allows us to investigate 
if these potential dynamical problems do arise or not. In particular, the 
confinement of constituent quarks --- out of reach of conventional methods ---
can be directly addressed with lattice methods. These investigations will
allow one to decide if the chiral quark model indeed provides a satisfactory 
explanation for the success of the non-relativistic quark model. Along the way,
one will be able to sharpen the algorithmic tools of dynamical fermion 
simulations in a model that shares many features with QCD --- most important a 
manifest chiral symmetry --- but which should be much simpler to simulate than 
Ginsparg-Wilson fermions in the chiral limit.

At this stage, it is not clear if lattice theories with a nonlinearly realized
chiral symmetry can also provide a formulation of full QCD beyond low-energy 
effective theories. In order to answer this question, one may attempt to relate
lattice QCD with Ginsparg-Wilson fermions to a lattice theory with explicit
pion fields and a nonlinearly realized chiral symmetry. If such a connection
can be found analytically, one would expect that the anomalies of 
Ginsparg-Wilson fermions manifest themselves as Wess-Zumino-Witten and 
Goldstone-Wilczek terms of the lattice pion field. This is an interesting 
project for future studies which holds the promise to further deepen our 
understanding of the non-perturbative regularization of chiral symmetry.

\section*{Acknowledgements}

We would like to thank R.~C.~Brower, G.~Colangelo, J.~Gasser, P.~Hasenfratz, 
J.~Jersak, H.~Leutwyler, M.~L\"uscher, A.~Manohar, and F.~Niedermayer for 
helpful discussions. This work was supported in part by funds provided by the 
U.S.\ Department of Energy (D.O.E.) under cooperative research agreement 
DOE-FG02-96ER40945, by the Schweizerischer Nationalfond (SNF), by the European 
Community's Human Potential Programme under contract HPRN-CT-2000-00145, and by
the German Academic Exchange Service (DAAD).

\end{document}